%% file: learnedftl.tex
\documentclass[sigconf,screen,nonacm]{acmart}

\usepackage[normalem]{ulem}

\usepackage{graphicx}
\usepackage{float}
\usepackage{rotating}
\usepackage{comment}
\usepackage{algorithmicx}
\usepackage{algpseudocode}
\usepackage{makecell}
\usepackage{subcaption}
\usepackage{balance}
\usepackage{def}
\usepackage{tikz}
\usepackage{soul}
\sethlcolor{yellow}
\usepackage{textcomp}
\usepackage{enumitem}
\usepackage{adjustbox}
\usepackage[ruled,linesnumbered]{algorithm2e} 

\AtBeginDocument{%
  }

\setcopyright{acmcopyright}
\acmConference[ASPLOS '23]{Architectural Support for Programming Languages and Operating Systems}
{March 25--29, 2023}{Vancouver, Canada}
\acmBooktitle{Proceedings of Architectural Support for Programming Languages and Operating Systems (ASPLOS ’23), March 25--29, 2023, Vancouver, Canada}
\copyrightyear{2023}
\acmYear{2023}
\acmDOI{XXXXXXX.XXXXXXX}

\begin{document}

\title{\pname{}: A Learning-based Flash Translation Layer \\for Solid-State Drives}


\author{Jinghan Sun}
\affiliation{%
\institution{UIUC} \country{}}
\email{js39@illinois.edu}

\author{Shaobo Li}
\affiliation{%
\institution{UIUC} \country{}}
\email{shaobol2@illinois.edu}

\author{Yunxin Sun}
\affiliation{%
  \institution{ETH Zurich}
  \country{}
}
\email{yunsun@student.ethz.ch}

\authornote{Work done when visiting the Systems Platform Research Group at UIUC as a research intern.}

\author{Chao Sun}
\affiliation{%
  \institution{Western Digital Research}
  \country{}
}
\email{chao.sun@wdc.com}

\author{Dejan Vucinic}
\affiliation{%
  \institution{Western Digital Research}
  \country{}
}
\email{dejan.vucinic@wdc.com}

\author{Jian Huang}
\affiliation{%
\institution{UIUC} \country{}}
\email{jianh@illinois.edu}

\renewcommand{\shortauthors}{Jinghan Sun, Shaobo Li, Yunxin Sun, Chao Sun, Dejan Vucinic, and Jian Huang}
\renewcommand{\shorttitle}{LeaFTL: A Learning-based Flash-Translation Layer for Solid-State Drives}

\input{abstract}





\maketitle

\input{intro}

\input{motivation}
\input{design}

\input{eval}

\input{discussion}
\input{related}

\input{conclusion}

\input{ack}



\balance
\bibliographystyle{ACM-Reference-Format}
\bibliography{ref}


\end{document}

%% file: abstract.tex
\begin{abstract}
In modern solid-state drives (SSDs), the indexing of flash pages is a critical component in their storage controllers.  
It not only affects the data access performance, but also determines the efficiency of the precious in-device DRAM resource. 
A variety of address mapping schemes and optimizations have been proposed. However, most of them were developed 
with human-driven heuristics. 

In this paper, we present a learning-based flash translation layer (FTL), named \pname{}, which learns the address mapping to 
tolerate dynamic data access patterns via linear regression at runtime. By grouping a large set of mapping entries into a 
learned segment, it significantly reduces the memory footprint of the address mapping table, which further benefits the data 
caching in SSD controllers. \pname{} also employs various optimization techniques, including out-of-band metadata verification 
to tolerate mispredictions, optimized flash allocation, and dynamic compaction of learned index segments. We implement \pname{} 
with {both a validated SSD simulator and a real open-channel SSD board. Our evaluation with various storage workloads demonstrates that  
\mbox{\pname{}} saves the memory consumption of the mapping table by 2.9$\times$ and improves the storage performance 
by 1.4$\times$ on average, in comparison with state-of-the-art FTL schemes. }

\end{abstract}

%% file: intro.tex
\begin{figure*}[t]
\centering
\includegraphics[scale=0.82]{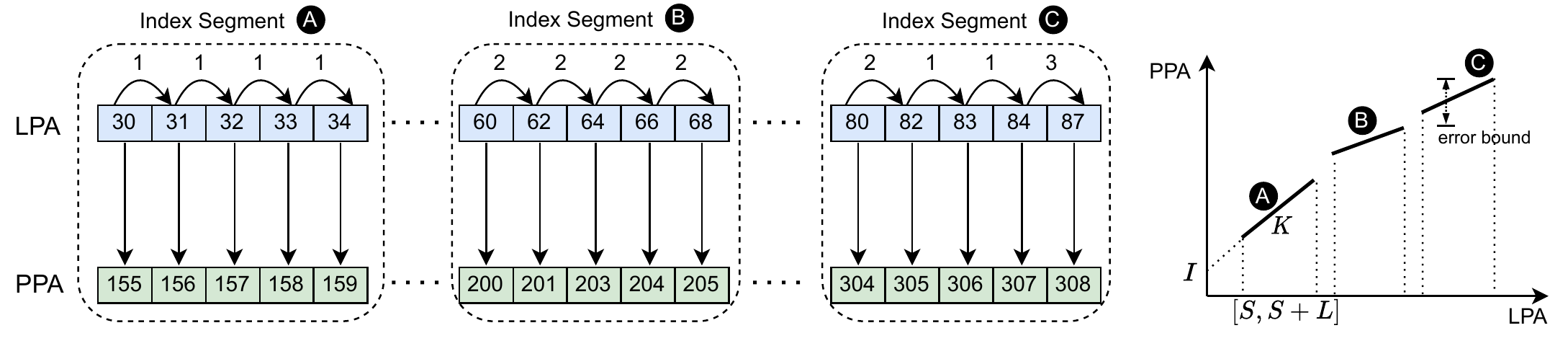}
\vspace{-3ex}
\caption{An illustrative example of learning LPA-PPA mappings using piecewise linear regression in \pname{}. 
It can learn various patterns of LPA-PPA mappings with guaranteed error bound. Each learned index segment can be 
represented with $(S, L, K, I)$, where $[S, S+L]$ denotes the interval of LPAs, $K$ is the slope, and $I$ is the intercept of the index segment. 
}
\label{fig:casestudy}
\end{figure*}

\section{Introduction}
\label{sec:intro}
Flash-based SSDs have become an indispensable part in modern storage systems, 
as they outperform conventional hard-disk drives (HDDs) by orders of magnitude, 
and their cost is close to that of HDDs~\cite{flashblox:fast2017, rssd:asplos2022, iceclave:micro2021, timessd:eurosys2019}. 
The SSD capacity continues to boost 
by increasing the number of flash channels and chips with the rapidly shrinking process and 
manufacturing technology~\cite{ouyang:asplos2014, sftl:msst2011, flashblox:fast2017, deepstore:micro2019}.

The flash translation layer (FTL) is the core component of managing 
flash memory in SSDs, including address translation,  
garbage collection (GC), and wear leveling~\cite{DFTL, zhang:fast2012}. 
The FTL maintains metadata structures for different functions such as address translation and valid page tracking, 
and caches them in the in-device DRAM (SSD DRAM) for improved performance~\cite{sftl:msst2011, dayan:sigmod2016, chen:fast2011}.

Among these data structures, the address mapping table has the largest memory footprint. 
In general, the address mapping table can be categorized in three types: page-level mapping, 
block-level mapping, and hybrid mapping. Modern SSDs usually use the page-level mapping, as it offers 
the best performance for the flash page lookup, and incurs minimal GC overhead, in comparison with 
the other two mapping schemes~\cite{DFTL, zhang:fast2012}. However, the page-level mapping table size  
is large, as it stores the entry for the LPA-to-PPA address translation for each flash page. 


The address mapping table significantly affects the performance of SSDs, as it not only  
determines the efficiency of indexing flash pages, but also affects the utilization of 
SSD DRAM. Moreover, due to the limitations of the cost and power budget in SSD controllers, 
it is challenging for SSD vendors to scale the in-device DRAM capacity~\cite{dayan:sigmod2016, 
deepstore:micro2019}. This challenge becomes even worse 
with the increasing flash memory capacity in an SSD, as larger capacity usually requires a larger
address mapping table for indexing. 

To improve the address mapping and translation for SSDs, various optimization schemes
have been developed~\cite{sftl:msst2011, zhang:fast2012, chung:sac2011, kang:emsoft2006, 
lee:osr2008, lee:toecs2007}. However, most of them were developed based on human-driven heuristics~\cite{sftl:msst2011}, 
and cannot capture dynamic data access patterns at runtime.
Employing more semantic knowledge into the FTL, such as GraphSSD~\cite{graphssd:isca2019}, can improve the data indexing 
and address translation, however, it is application specific and complicates 
the management of address mappings~\cite{chen:fast2011}, 
which does not scale for the development of generic SSDs. In this work, we do not expect that 
we can obtain application semantics from the host and the SSD controller. 
Instead, we focus on utilizing {simple yet effective} machine learning (ML) techniques to automate the address mapping table 
management in the SSDs, with the capability of learning diverse and dynamic data access patterns.

To this end, we propose a learning-based FTL, named \pname{}, by
utilizing the piecewise linear regression technique to learn the LPA-PPA mappings, and automatically exploiting
the data locality of various data access patterns at runtime. Unlike the state-of-the-art page-level mapping, 
the key idea of \pname{} is that it can learn the correlation 
between a set of LPAs and their mapped PPAs, based on which it can build a space-efficient index segment, 
as presented in \circleb{A} in Figure~\ref{fig:casestudy}. Since the learned 
index segment can be simply represented with $(S, L, K, I)$, where $[S, S+L]$ denotes the interval of LPAs, $K$ is 
the slope of the segment, and $I$ is the intercept {of the segment (see the last diagram in Figure~\mbox{\ref{fig:casestudy}})}, 
each segment will take only 8 bytes 
(1 byte for $S$ and $L$, 2 bytes for $K$, and 4 bytes for $I$) with our optimizations (see the details 
in $\S$\ref{sec:design}).  
Compared to the on-demand page-level mapping~\cite{DFTL}, the learned segment reduces the mapping table 
size by a factor of $m*avg(L)/8$, where $m$ is the size (8 bytes) of each entry in the on-demand 
page-level mapping table, and $avg(L)$ is the average number of LPA-PPA mappings that can be represented in a 
learned index segment, $avg(L)$ is 20.3 according to our study of various storage workloads. 

Beyond learning contiguous LPA-PPA mappings, \pname{} also learns different correlation patterns, 
such as regular and irregular strided data accesses as shown in \circleb{B} and \circleb{C}, respectively. 
Unlike existing indexing optimizations based on human-driven heuristics, \pname{} can learn more 
irregular patterns of LPA-PPA mappings with guaranteed error bound, 
as shown in \circleb{C}. This enables \pname{} to further condense the address mapping table. 
Therefore, given a limited DRAM capacity in the SSD controller, \pname{} can maximally utilize 
the DRAM caching and improve the storage performance.  
For the worst case like random I/O accesses, \pname{} will transfer the mapping into
single-point linear segments ($L=0$, $K=0$, and $I=PPA$ in Figure~\ref{fig:casestudy}),
and its memory consumption will be no more than that of the page-level mapping.

With the learned index segments, \pname{} may occasionally return an 
inaccurate PPA (i.e., address misprediction), which incurs additional flash accesses until 
the correct PPA is identified. To overcome this challenge, we develop an 
error-tolerant mechanism in \pname{}. For each flash page access, we use the reverse 
mapping stored in the out-of-band (OOB) metadata of each flash page to verify the correctness of 
the data access. 
Since the OOB usually has 64--256 bytes{~\mbox{\cite{DFTL, FlashMap}}}, we use it to store the accurate LPAs mapped to the 
neighbor PPAs. Thus, upon an address misprediction, we use the stored reverse mappings to 
find the correct PPA, avoiding additional flash accesses. {\pname{} leverages the intrinsic OOB structure 
to handle address mispredictions and make SSD perfectly-suited for practical learned indexing.}


Due to the intrinsic out-of-place write property of SSDs (see $\S$\ref{sec:motivation}), the learned index segments will be disrupted by  
writes and GC, and the segments need to be relearned with new LPA-PPA mappings. 
To tolerate these disruptions, 
the learned segments are organized within multiple levels to maintain the temporal order in a log-structured manner:
the topmost level has the most recent segments, and the lower level stores older segments.
The segments at the same level are sorted without overlapping.
If the new segment has a conflict with an existing segment, the old segment will be moved to
the lower level. 
Therefore, \pname{} can always identify the latest version of the corresponding LPA-PPA mapping 
in a top level of learned index segments. 
\pname{} will compact the learned segments periodically to reduce its memory footprint. 

To further maximize the efficiency of \pname{}, we coordinate its learning procedure with 
flash block allocation in the SSD. As flash block allocation decides the distribution of 
mapped PPAs, \pname{} will allocate consecutive PPAs to contiguous LPAs at its best effort, for 
increasing the possibility of learning a space-efficient index segment. Similar to existing  
page-level mapping~\cite{DFTL, FlashMap}, \pname{} stores the learned index segments in 
flash blocks for recovery. 
Overall, we make the following contributions:

\begin{itemize}[leftmargin=*]

\vspace{1ex}
\item We present a learning-based FTL, 
it can learn various data access patterns and turn them into index segments  
for reducing the storage cost of the mapping table.  

\vspace{1ex}
\item We develop an error-tolerant address translation mechanism to handle address 
mispredictions caused by the learned indexes, with minimal extra flash accesses. 

\vspace{1ex}
\item We preserve the core FTL functions, and enable the coordination between the 
learning procedure of the address mapping table with the flash block allocation and 
GC to maximize the efficiency of the learned FTL.  

\vspace{1ex}
\item We manage the learned segments in an optimized log-structured manner, and enable 
	compaction to further improve the space efficiency for the address mapping.    

\vspace{1ex}
\end{itemize}

We implement \pname{} with {a validated SSD simulator WiscSim~\mbox{\cite{wiscsim:eurosys2017}}  
and evaluate its efficiency with a variety of popular storage workloads. We also develop a system prototype 
with a real 1TB open-channel SSD to verify the functions of \mbox{\pname{}} and 
validate its efficiency with real data-intensive applications, such as the key-value store and transactional database. 
Our evaluation with the real SSD shows 
similar benefits as that of the SSD simulator implementation. We demonstrate that \pname{}} reduces 
the storage cost of the address mapping in the FTL by 2.9$\times$ on average. The saved memory space 
benefits the utilization of the precious SSD DRAM, and further improves the storage performance 
by 1.4$\times$ on average.
We also show that \mbox{\pname{}} does not affect the SSD lifetime, and its learning procedure introduces negligible performance 
overhead to the storage processor in the SSD controllers. 
The codebase of \pname{} is available at \url{https://github.com/platformxlab/LeaFTL}.



%% file: motivation.tex
\section{Background and Motivation}
\label{sec:motivation}

\begin{figure}
\centering
\includegraphics[width=0.75\linewidth]{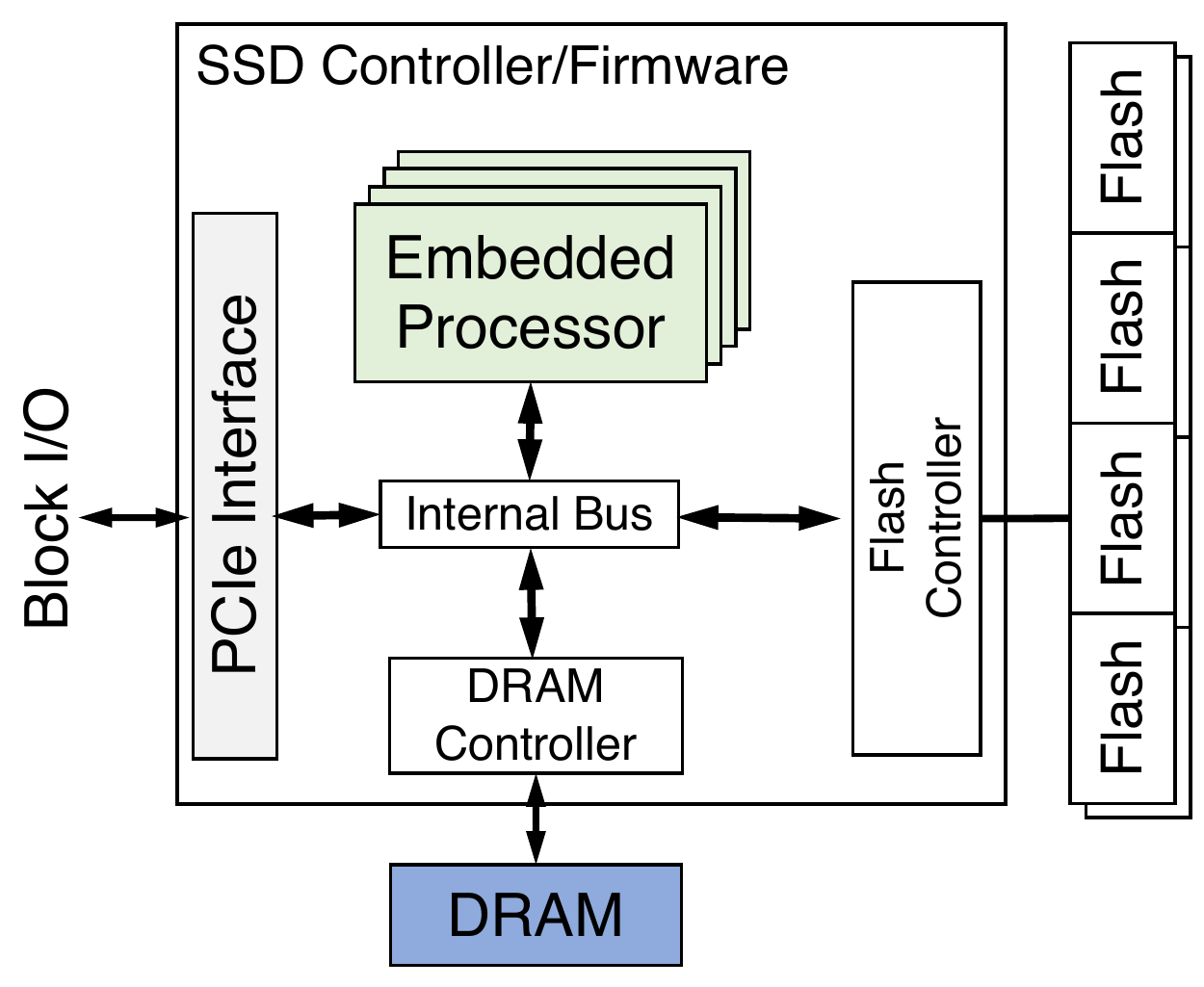}
\vspace{-1ex}
\caption{The internal system architecture of SSDs.}
\vspace{-3ex}

\label{fig:ssd-arch}
\end{figure}

\noindent
\textbf{Flash-Based Solid-State Drive.}
An SSD has three major parts (see Figure~\ref{fig:ssd-arch}): a set of flash memory packages, 
an SSD controller with embedded processors, and a set of flash controllers.
With the nature of NAND Flash,  
when a free page is written, the page cannot be written again until that
page is erased. However, erase operation is performed only
at a block 
granularity. As the erase operation is expensive, writes are issued to free flash pages 
erased in advance (i.e., out-of-place write).
GC will be performed to clean
the stale data. As each flash block has limited endurance, 
it is important for them to age uniformly (i.e., wear leveling).
SSDs have a logical-to-physical address mapping table to index flash pages. 
All these functions are managed by the FTL in the SSD firmware.

Modern SSD controllers have 
general-purpose embedded processors {(e.g., ARM processors)}. 
The processors help with issuing I/O requests,
translating LPAs to PPAs, and handling GC and wear-leveling.
SSDs also have limited DRAM capacities to cache the mapping table and the application data. 

\noindent
\textbf{Address Mapping Table in the FTL.}
The address mapping table in FTL generally has three types: page-level mapping, block-level mapping, 
and hybrid mapping. The page-level mapping enables direct LPA-PPA mapping for fast lookup. 
However, each entry usually takes 8 bytes (4 bytes for LPA, 4 bytes for PPA), and the entire 
mapping table requires large storage space. 
The block-level mapping significantly reduces the mapping table size. However, it introduces
additional overhead for the page lookup in the flash block. The hybrid mapping takes 
advantages of both page-level and block-level mapping. It uses 
log blocks to store new writes, and index them with the page-level mapping. The log blocks will be 
moved into data blocks that are indexed with block-level mapping. This incurs significant GC overhead. Therefore, modern SSDs commonly use the 
page-level mapping scheme. 

\begin{figure}
\centering
\includegraphics[width=\linewidth]{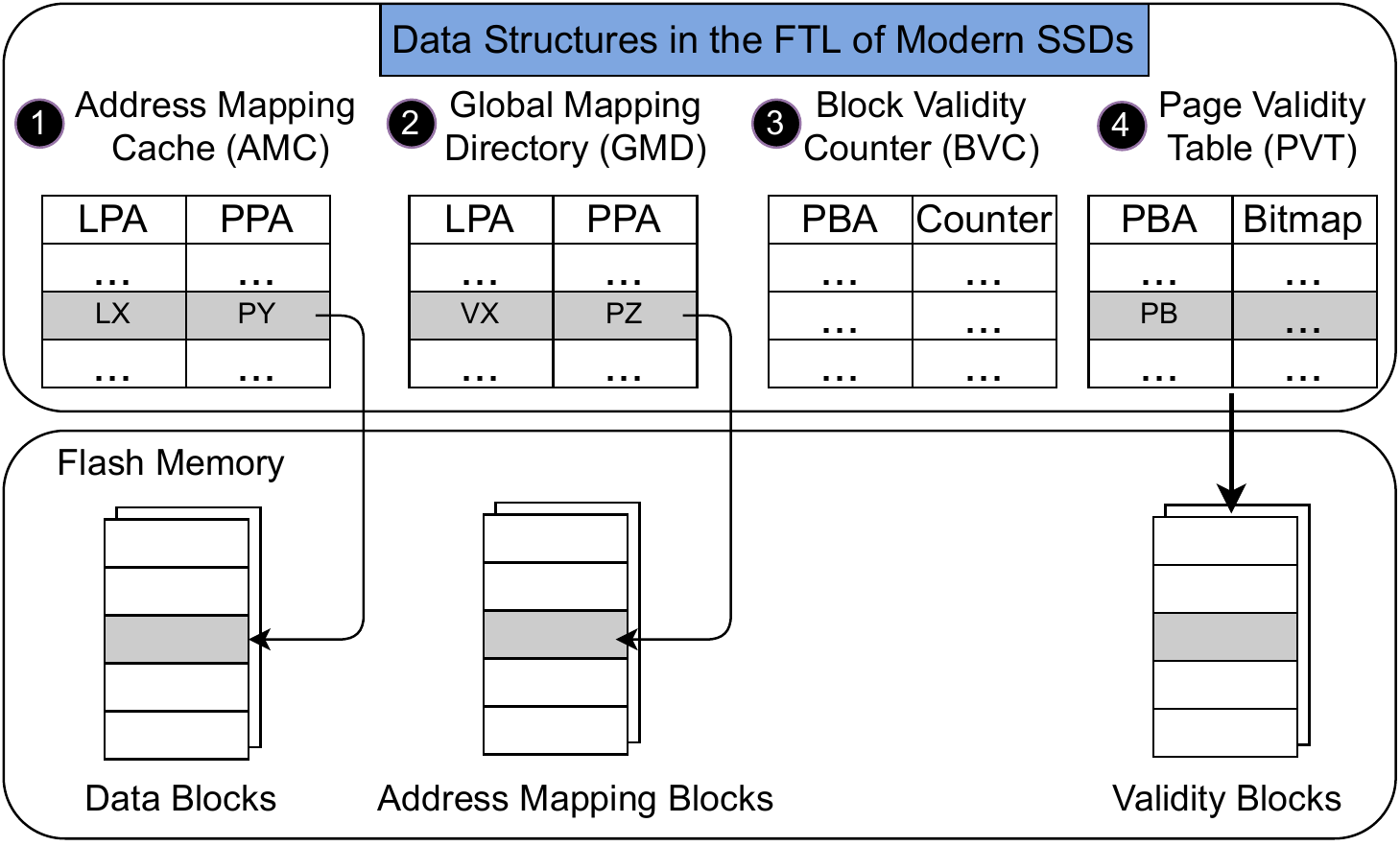}
\caption{The common data structures in the FTL of SSDs.}
\vspace{-4.5ex}
\label{fig:ftl-structure}
\end{figure}

\noindent
\textbf{Metadata Structures for Flash Management.}
The FTL usually employs four metadata structures (see Figure~\ref{fig:ftl-structure}): (1) the address mapping cache (\circleb{1} AMC) for 
caching the address mapping table in the SSD DRAM; (2) the global mapping directory (\circleb{2} 
GMD) for tracking the locations of the address mapping table pages in the SSD; 
(3) the block validity counter (\circleb{3} BVC) for tracking 
the number of valid pages for each flash block for assisting the GC in the SSD; and (4) the page validity 
table (\circleb{4} PVT), which uses bitmaps to track the valid pages in each flash block. 
During the GC, the FTL will check the \circleb{3} BVC to select candidate flash blocks, 
and migrate their valid pages to free flash blocks. After that, 
it will erase these selected flash blocks, and mark them as free blocks.


\noindent
\textbf{Limited DRAM Capacity in SSD Controllers.}
It is hard to provision large DRAM inside SSD controllers, due to their   
hardware constraints and limited budgets for power and hardware cost~\cite{deepstore:micro2019, dayan:sigmod2016, ssddram}.  
Thus, SSD controllers often use on-demand caching to maintain the recently accessed metadata and data 
in the SSD DRAM. 

Among all the metadata structures, the address mapping table has the largest memory footprint. 
As discussed, \circleb{1} AMC caches the recently accessed mapping table entries.
If a mapping entry is not cached, the FTL will locate the
corresponding address mapping table pages stored in the flash blocks, and place the mapping entry in the \circleb{1} AMC. 
As we scale the SSD capacity, the DRAM challenge will become even worse.   
To overcome this challenge, various optimizations on the mapping table have been 
proposed~\cite{sftl:msst2011, chung:sac2011, kang:emsoft2006, lee:osr2008, lee:toecs2007, bast} to improve the utilization of the SSD DRAM. 
However, most of them cannot automatically capture 
diverse data access patterns at runtime, leaving a large room for improvement. 

%% file: design.tex
\section{Design and Implementation}
\label{sec:design}



To develop \pname{} 
in the SSD controller, we have to overcome the following research challenges. 

\begin{itemize}[leftmargin=*]

\vspace{1ex}
\item \mbox{\pname{}} should be able to automatically capture diverse data access 
patterns, and generate memory-efficient address mapping (\mbox{$\S$\ref{subsec:keyidea}, 
$\S$\ref{subsec:learnedsegment}, $\S$\ref{subsec:learnedoptimization}, and $\S$\ref{subsec:learnedmapping}}).

\vspace{1ex}
\item \pname{} may incur address mispredictions, which could incur additional flash 
accesses. \pname{} should be tolerant of errors and have low misprediction penalty ($\S$\ref{subsec:error}).  

\vspace{1ex}
\item \pname{} should work coordinately with other core FTL functions that include 
GC and wear leveling ($\S$\ref{subsec:gc}).  

\vspace{1ex}
\item \pname{} should be lightweight and not incur much extra overhead 
to storage operations 
($\S$\ref{subsec:operation}, $\S$\ref{subsec:workflow} and $\S$\ref{subsec:implt}).

\end{itemize}

\begin{figure}
    \centering
    \includegraphics[width=0.4\textwidth]{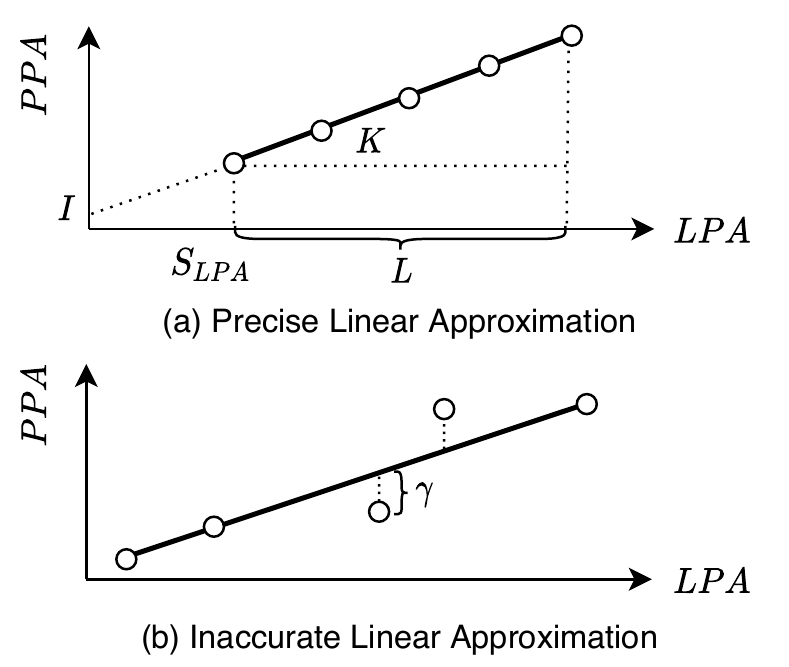}
    \vspace{-1ex}
    \caption{Visualization of learned index segments.}
    \vspace{-1ex}
    \label{fig:keyidea}
\end{figure}

\subsection{Key Ideas of \pname{}}
\label{subsec:keyidea}

Instead of using the space-consuming one-to-one mapping in the page-level mapping, 
the key idea of \pname{} is to exploit learning techniques to identify various LPA-PPA mapping patterns  
and build efficient learned address mapping entries. 
{
Modern SSD controllers usually have a data buffer for 
grouping writes and write the large data chunk at once for exploiting the internal flash parallelisms.  
\mbox{\pname{}} utilizes this data buffer to collect LPA-to-PPA mappings for learning index segments for free, and 
does not introduce extra data collection overhead (see the details in $\S$\mbox{\ref{subsec:learnedoptimization}}).}

As shown in Figure~\ref{fig:keyidea} (a), the PPA of an LPA can be obtained with the expression: 
$PPA = f(LPA) = \lceil K*LPA + I \rceil$, $LPA \in [S_{LPA}, S_{LPA}+L]$, where $[S_{LPA}, S_{LPA}+L]$ denotes the 
interval ($L$) of LPAs, $K$ is the slope, and $I$ is the intercept. 
As discussed in $\S$\ref{sec:intro}, each learned index segment can be represented in 8 bytes: 1 byte for $S_{LPA}$ and $L$, respectively; 
2 bytes for $K$, and 4 bytes for $I$. The size of $S_{LPA}$ is reduced from 4 bytes to 1 byte with 
our optimizations on the segment management (see $\S$\ref{subsec:learnedmapping}).  


\begin{figure}[t]
	\centering
	\includegraphics[width=0.98\linewidth]{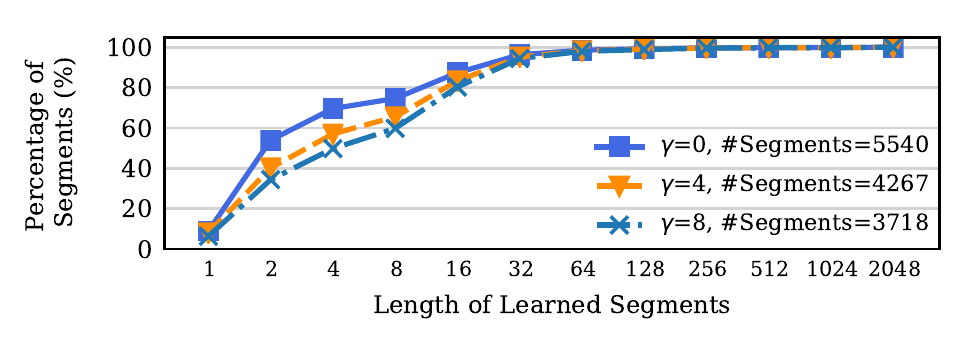}
	\vspace{-2ex}
        \caption{Aggregated distribution of learned segments.}
	\label{fig:segment}
    \vspace{-1ex}
\end{figure}

We can relax the linear regression to capture more flash access patterns, which further reduces the 
learned address mapping table size. As shown in Figure~\ref{fig:keyidea} (b), the linear regression can 
learn a pattern with guaranteed error bound $[-\gamma, \gamma]$. As we increase $\gamma$, we can cover more flash access patterns. 
We applied the relaxed linear regression with different $\gamma$ values to a variety of storage workloads (see 
$\S$\ref{subsec:setup}), our 
experimental results demonstrate that the number of learned index 
segments is gradually decreased, as we increase $\gamma$. Figure~\ref{fig:segment} shows that 
98.2--99.2\% of the learned index segments cover up to 128 LPA-PPA mapping entries, demonstrating the potential 
advantages of the learning-based approach. 

As for random access patterns, 
\pname{} will transfer the learned segments into single-point segments. 
And these linear segments do not require more storage space than the page-level mapping.

\begin{figure}[t]
    \centering
    \includegraphics[width=0.46\textwidth]{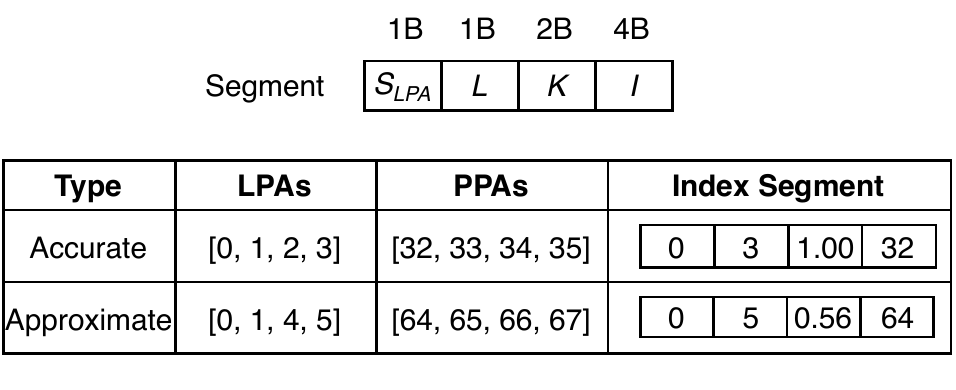}
	\vspace{-1ex}
	\caption{Types of learned segments in \pname{}.}
    \label{fig:types}
\end{figure}

\subsection{Learned Index Segment}
\label{subsec:learnedsegment}

%
%


\paragraph{Types of Learned Index Segment.}
The mapping table of \pname{} is built with learned index segments.
It has two types of segments: accurate and approximate segments, 
as shown in Figure~\ref{fig:types}. {Both of them are learned with piecewise linear regression technique~\mbox{\cite{pgmindex:vldb2014}}}. 

As for the accurate index segments, given an LPA, we can precisely get the corresponding PPA with $f(LPA) = \lceil K*LPA + I \rceil$. For example, when the LPA is 2 in Figure~\ref{fig:types}, 
we can directly get the PPA value of 34 with $\lceil1.00*2 + 32\rceil$. In this example,
the learned segment has $L = 3$ and it indexes 4 LPA-PPA mappings. If $L = 0$, the learned segment will 
become a single-point segment, the slope $K = 0$, and we will get its PPA with $PPA = I$.  

As for approximate index segments, we use the same formula $f(LPA) = \lceil K*LPA + I \rceil$ to calculate 
the PPA. However, the returned PPA may not be the exact corresponding PPA. It has an error bound 
$[-\gamma, \gamma]$ guaranteed by the linear regression, and $\gamma$ is configurable. 
For example, given $LPA = 4$ in Figure~\ref{fig:types}, 
the value of the PPA is 67, according to the calculation $\lceil4*0.56 + 64\rceil$. However, the 
real PPA should be 66. We define this as \textit{address misprediction}. We will discuss how we handle the address misprediction with 
reduced miss penalty in $\S$\ref{subsec:error}. 

\begin{figure}[t]
    \centering
    \includegraphics[width=0.38\textwidth]{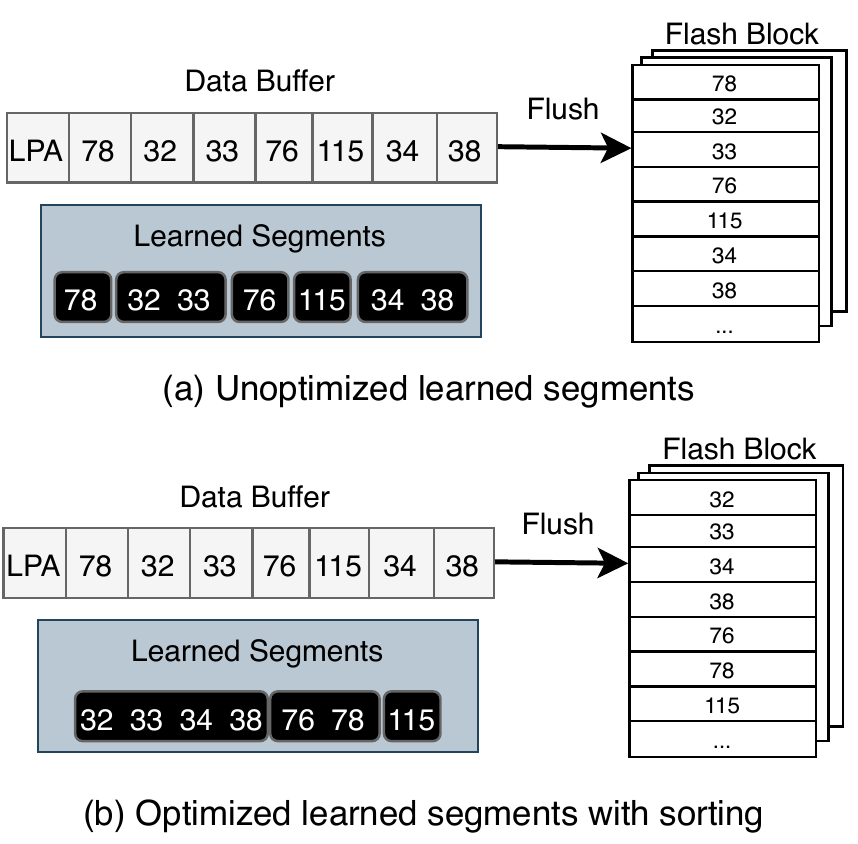}
	\vspace{-1ex}
    \caption{An example of reducing the number of learned segments via exploiting the flash block allocation.}
    \label{fig:sorting}
    \vspace{-3ex}
\end{figure}

\noindent
\paragraph{Size of Learned Index Segment.}
As discussed in $\S$\ref{subsec:keyidea}, each segment can be expressed in $(S_{LPA}, L, K, I)$.  
The starting LPA will take 4 bytes.  
We can further reduce this size by partitioning a range of LPAs into small groups, and each LPA group represents a 
certain number of contiguous LPAs. Therefore, we can index an LPA with its offset in a corresponding group. 
In \pname{}, each group represents 256 contiguous LPAs. 
Thus, $S_{LPA}$ can be indexed by the offset ($2^8 = 256$) in the group, which takes only 1 byte.  
We use 256 as the group size, because the length of the learned segments is usually less than 
256 (see Figure~\ref{fig:segment}). 

Given an LPA, we can get its offset in the group with $(LPA\ mod\ $ $256)$.
In \pname{}, we set the $L$ as 1 byte. Thus, each segment can index 256 LPA-PPA mappings. We use a 
16-bit floating point to store the value of the slope $K$. And the intercept $I$ of a segment can be 
represented in 4 bytes. Therefore, in combination with $S_{LPA}$, both accurate and approximate segments 
can be encoded with 8 bytes (see Figure~\ref{fig:types}), which are memory aligned. 

\pname{} uses the least significant bit of the $K$ to indicate segment 
types (0 for accurate segments, 1 for approximate segments). This has negligible impact on the 
address translation accuracy, because $K \in [0, 1]$, which will only affect the tenth digit after decimal point. 



\subsection{Improve the Learning Efficiency}
\label{subsec:learnedoptimization}

To further reduce the number of learned segments, \pname{} performs optimizations to 
improve its learning efficiency of address mappings by exploiting the flash block allocation in SSD controllers, as 
shown in Figure~\ref{fig:sorting}. Flash pages are usually buffered in the SSD controller 
and written to flash chips at a flash block granularity, for utilizing the internal bandwidth and avoiding the 
open-block problem~\cite{open-block-issue, flashblox:fast2017, lee2016application, cai2017error}. This allows \pname{} to learn more space-efficient index segments (i.e., index segments can cover more LPA-PPA mappings) 
by reordering the flash pages with their LPAs in the data buffer. 
As shown in Figure~\ref{fig:sorting} (a), \pname{} learns 5 index segments (78), (32, 33), (76), (115), and (34, 38) with 
$\gamma = 4$. After sorting the pages in the data buffer shown in Figure~\ref{fig:sorting} (b), \pname{} generates  
3 index segments (32, 33, 34, 38), (76, 78), and (115). 

To develop the optimized learned segments, \pname{} sorts the flash pages in ascending order of their LPAs in the 
data buffer (8MB by default). 
When pages in the data buffer is flushed to the flash chips, 
their PPAs are in ascending order. This ensures a monotonic address mapping 
between LPAs and PPAs, which reduces the number of index segments. 

\subsection{Manage Learned Index Segments}
\label{subsec:learnedmapping}
Upon new data updates or GC in the SSD, the learned index segments 
need to be updated, due to the intrinsic property (i.e., out-of-place update) of SSDs. 
Unfortunately, the direct updates to learned index segments are expensive, since we have to relearn 
the index segments with new PPAs. {This relearning procedure not only consumes extra compute cycles,  
but also involves additional flash accesses, since we have to access the corresponding flash pages to 
obtain accurate PPAs for some of the LPAs in the index segment being updated. For instance, for in-place update to an approximate segment, 
it can incur 21 flash accesses on average when relearning. In-place update also breaks the existing LPA-to-PPA mapping patterns, which results in $1.2\times$ additional segments and memory footprint, according to our experiments with 
various workloads.}

\begin{figure}[t]
    \centering
    \includegraphics[width=0.98\linewidth]{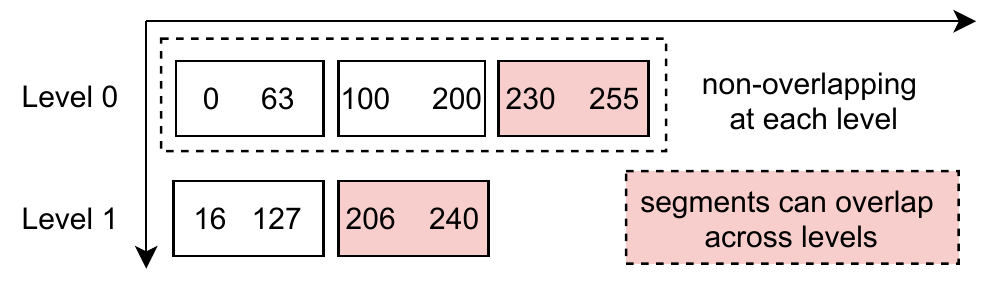}
	\caption{The learned index segments are managed in a log-structured manner in \pname{}.}
    \label{fig:hierarchy}
\end{figure}

To address this challenge, we manage the learned index segments in a log-structured manner, as shown 
in Figure~\ref{fig:hierarchy}. Therefore, the newly learned index segments will be appended to the log structure 
(level 0 in Figure~\ref{fig:hierarchy}) and used to 
index the updated LPA-PPA mappings, while the existing learned segments (level 1 and lower levels in Figure~\ref{fig:hierarchy}) 
can still serve address translations for LPAs whose mappings have not been updated. 
{Such a structure supports concurrent lookups as enabled in the traditional log-structured merge tree. As we insert the 
newly learned index segments at the top level of the log-structured tree, this minimizes the impact on other segments.}

\noindent
\paragraph{Log-Structured Mapping Table.}
The log-structured mapping table has multiple levels to maintain the 
temporal order of index segments. As discussed, the topmost level has the most recent learned index segments, 
and the lower level stores the older segments. For the segments on the same level, \pname{} ensures that 
they are sorted and do not have overlapped LPAs. This is for fast location of the corresponding learned index 
segments in each level. For the segments across the levels, they may have overlapped LPAs, due to the 
nature of the log-structured organization. And the segments with overlapped LPA-PPA mappings will be compacted periodically  
for space reclamation (see its detailed procedure in $\S$\ref{subsec:operation}).

\noindent
\paragraph{Manage Two Types of Index Segments.}
\pname{} manages the accurate and approximate index segments in the same log-structured mapping table, 
as they can be encoded in the same format. For each accurate segment, we can directly infer its indexed 
LPAs with the $S_{LPA}$, $K$, and $L$, since it has a regular pattern. 
However, for approximate index segments, we only have the knowledge of the starting LPA and 
the end LPA with $S_{LPA} + L$. Its encoded LPAs cannot be directly inferred from their  
metadata $(S_{LPA}, L, K, I)$, since they are learned from 
irregular access patterns and may have mispredictions.  

\begin{figure}[t]
    \centering
    \subfloat[Approximate index segments that index overlapped LPAs.]{
        \centering
        \includegraphics[width=.95\linewidth]{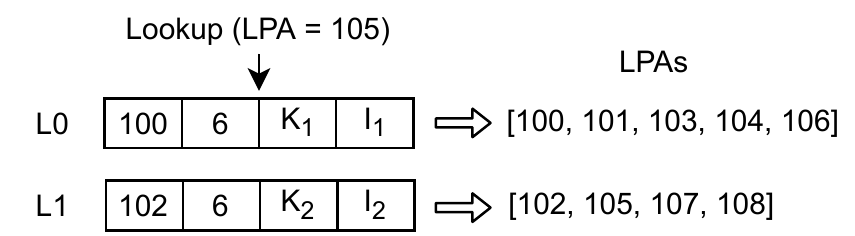}
    }
    \hspace{0.5cm}
    \subfloat[Resolve the conflict between approximate segments with CRB]{
        \centering
        \includegraphics[width=.95\linewidth]{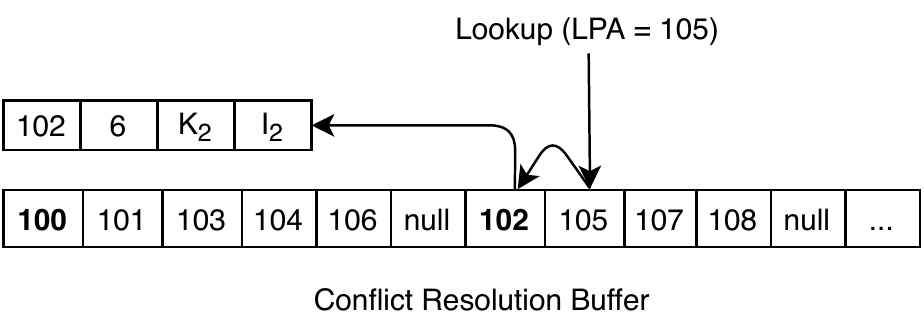}
    }
    \caption{A case study of conflict resolution buffer for approximate learned index segments.}
    	\vspace{-3ex}
    \label{fig:crb}
\end{figure}

If two approximate segments have overlapping LPA ranges, we could obtain inaccurate PPAs from the learned index segments. 
As shown in Figure~\ref{fig:crb} (a), given an LPA with the value 105, we will check the segment at Level 0 and 
may get an inaccurate PPA. 
This will also affect the efficiency of the segment compaction, with which we eliminate 
duplicated entries between segments. 


To address this challenge, \pname{} uses a Conflict Resolution Buffer (CRB) for each LPA group to store 
the LPAs indexed by each approximate segment. The main purpose of CRB is to help \pname{} check whether 
a given LPA belongs to one approximate segment. 

The CRB is a nearly-sorted list~\cite{nearly_sorted} 
by the starting LPAs of its approximate segments. 
To be specific, the CRB ensures the following properties: 
(1) the LPAs belong to the same approximate segment are stored contiguously; (2) different approximate 
segments are sorted by their starting LPA, and CRB uses a $null$ byte to separate these segments;
(3) it does not have redundant LPAs, which means an LPA will appear at most once in the CRB. This is 
achieved by removing existing same LPAs when we insert new approximate segments into the CRB.

However, if the $S_{LPA}$ of a new approximate segment is the same as any starting LPAs that have been 
stored in the CRB, \pname{} will update the $S_{LPA}$ of the old segment with the adjacent LPA. 
Take Figure~\ref{fig:crb} (b) as an example, upon a new approximate segment with $S_{LPA} = 100$, 
we will update the $S_{LPA}$ of the existing segment to 101, and then insert the new segment into the CRB. 
In this case, \pname{} will ensure each approximate segment will have its unique $S_{LPA}$. This will facilitate the approximate LPA-PPA address translation with high accuracy confidence.    

\begin{figure}[t]
    \centering
    \includegraphics[width=0.48\textwidth]{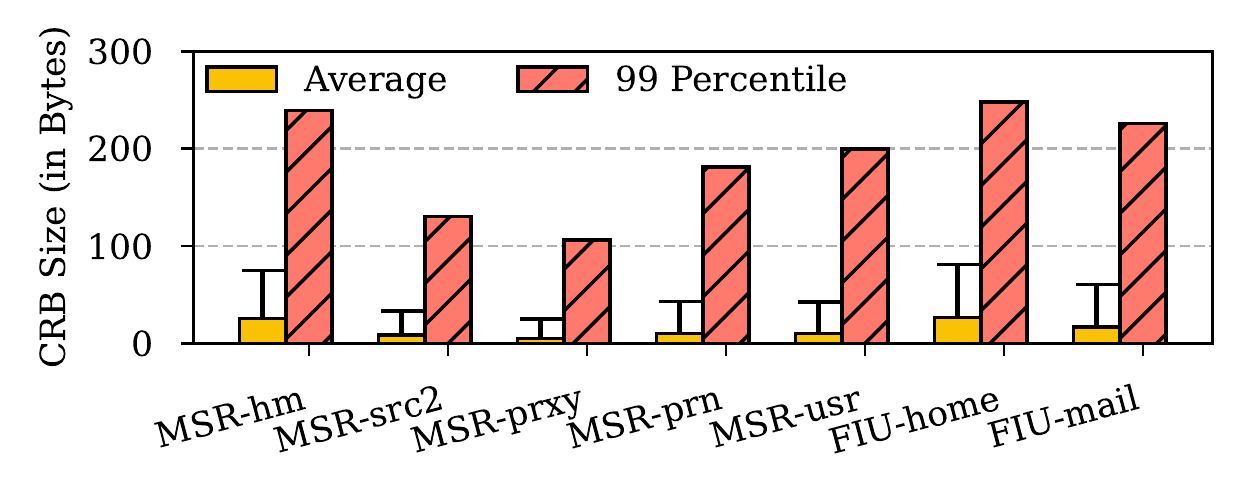}
     \vspace{-4ex}
	\caption{The distribution of CRB sizes for different storage workloads, when we set $\gamma = 4$ in \pname{}.}
    \vspace{-4ex}
    \label{fig:crbsize}
\end{figure}

Since CRB is nearly sorted, its insertion, deletion, and lookup operations are fast. 
The CRB is also space efficient, 
as each LPA (the offset in its corresponding LPA group) will take only one byte, and it guarantees that there are no 
redundant LPAs. Therefore, the CRB will maximally store 256 LPAs. Our experiments with a variety of storage 
workloads show that the CRB will take $13.9$ bytes on average, as shown in Figure~\ref{fig:crbsize}. 

Given an LPA, in order to identify which approximate index segment it belongs to, 
\pname{} will check the CRB with binary search. Once the LPA is found, \pname{} will search 
to its left until identifying the $S_{LPA}$, and this $S_{LPA}$ will be the starting LPA of the 
corresponding approximate segment, as shown in Figure~\ref{fig:crb} (b). Therefore, CRB can assist 
\pname{} to resolve the LPA lookups. 


\subsection{Handle Address Misprediction}
\label{subsec:error}
As discussed in $\S$\ref{subsec:learnedsegment}, the mapping table entries encoded with approximate segments may occasionally incur mispredictions 
and return an approximated PPA. 
These approximate segments have a guaranteed error bound $[-\gamma, \gamma]$, where $\gamma$ is a constant 
value that can be specified in the linear regression algorithm. 
To verify the correctness of the address translation, a simple method is to access the flash page with the predicted PPA, 
and use the reverse mapping (its corresponding LPA) stored in the OOB metadata of the 
flash page to check whether the LPA matches or not.  
In this case, upon a PPA misprediction, we need $\log(\gamma)$ flash accesses on average to 
identify the correct PPA. 

To avoid extra flash accesses for address mispredictions, \pname{} leverages the OOB of the flash page 
to store the reverse mappings of its neighbor PPAs. This is developed based on the insight that: with 
a $PPA_{learned}$ obtained from an approximate segment, its error bound $[-\gamma, \gamma]$ guarantees that the 
correct PPA is in the range of $[PPA_{learned}-\gamma, PPA_{learned}+\gamma]$, as discussed in Figure~\ref{fig:keyidea} (b). 
Thus, upon a misprediction, \pname{} 
will read the flash page with $PPA_{learned}$, and use its OOB to find the correct PPA. 
In this case, \pname{} ensures that it will incur only one extra flash access for address mispredictions.  

\begin{figure}[t]
    \centering
    \includegraphics[width=0.38\textwidth]{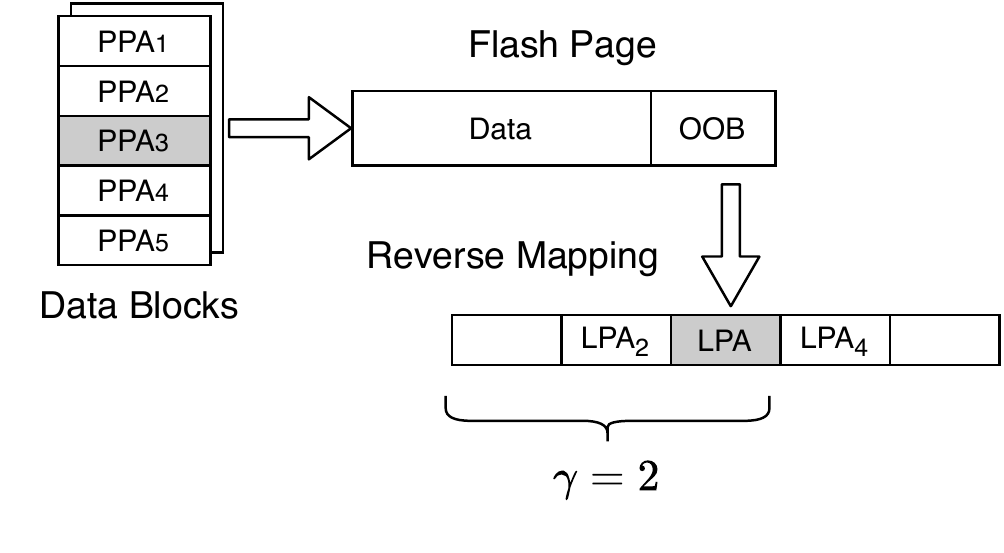}
	\vspace{-2ex}
    \caption{The out-of-band (OOB) metadata organization. It stores the reverse mapping for its neighbor PPAs.}
	\vspace{-2ex}
    \label{fig:oob}
\end{figure}

This is a feasible approach, as the OOB size is usually 128--256 bytes in modern SSDs. 
As each LPA takes 4 bytes, we can store 32--64 reverse mapping entries in the OOB. We show the OOB organization 
of \pname{} in Figure~\ref{fig:oob}. For the flash page $PPA_X$, the first 
$2\gamma + 1$ entries in its OOB correspond to the LPAs for the flash pages 
$[PPA_X - \gamma, PPA_X + \gamma]$. For the flash pages at the beginning and end of a flash block, 
we may not be able to obtain the reverse mapping of their neighbor PPAs. We place the $null$ bytes in the 
corresponding entry of the OOB.  

\subsection{Preserve Other Core FTL Functions}
\label{subsec:gc}
\pname{} preserves the core functions such as GC and wear leveling in an FTL. 
It follows the same GC and wear leveling policies in modern SSDs. 
When the number of free blocks in an SSD is below a threshold (usually 15-40\% of the total flash blocks), the SSD controller will trigger the GC execution.
\pname{} employs the greedy algorithm~\cite{nitin:atc2008} to select the candidate 
blocks which have the minimal number of valid pages, for reducing the data movement overhead at GC.  
As the GC move the valid pages from the candidate blocks to the free blocks, \pname{} places 
these valid pages into the DRAM buffer, sort them by their LPAs, and learn a new index segment.
The learning procedure is the same as we build index segments for new flash writes/updates. 
Thus, the address mapping of the valid pages is updated after the GC.

\pname{} also ensures all the flash blocks age at the same rate (i.e., wear leveling). 
It uses the throttling and swapping mechanism developed in existing GC, 
in which the cold data blocks (i.e., blocks not frequently accessed) will be migrated to hot blocks 
(i.e., blocks that experience more wear). 
\pname{} will learn new indexes for these swapped blocks and insert them into the mapping table 
to update their address mappings.



\subsection{\pname{} Operations}
\label{subsec:operation}
Now we describe the \pname{} operations, 
including segment {creation}, insert/update, LPA lookup, and compaction. We discuss 
their procedures, and use examples to illustrate each of them, respectively. 
We present their detailed procedures in Algorithm~\ref{algo:logplr} and ~\ref{algo:merge}.  

\newcommand\mycommfont[1]{\footnotesize\ttfamily\textcolor{purple}{#1}}
\SetCommentSty{mycommfont}
\SetAlFnt{\small\sffamily}
\begin{algorithm}[t]
    \DontPrintSemicolon 
    \SetAlgoNoEnd
    \SetAlgoLined
    \KwIn{$groups \leftarrow\ \pname{}\ group\  partitions$}
    \tcp{\color{purple}\textbf{Insert/Update Segment in the \pname{}}}
    \SetKwFunction{FMain}{$seg\_update$}
    \SetKwProg{Fn}{Function}{:}{}
    \Fn{\FMain{$segment, level$}}{

        $ seg\_pos = binary\_search(level, segment.S_{LPA})$ \;
        $ level.insert(segment, seg\_pos) $ \;
        {\If{$not\ segment.accurate$}{
            Insert LPAs into CRB and remove redundant LPAs \;
            {\If{$segment.S_{LPA}$ exists in CRB}{
                Update the $S_{LPA}$ of the old segment\;
                }
            }
        }}
        $ victim\_segments \leftarrow $ All segments that overlap the $segment$ starting with $seg\_pos$ \;
        {\ForEach{$victim \in victim\_segments$}{
            $ seg\_merge (segment, victim) $ \;
            \tcp*[l]{if marked as removable by seg\_merge()}
            {\If{$victim.L = -1$ }
                {
                $level.remove(victim)$ }
            }
            {\If{$segment.overlaps(victim)$}
                {Pop $victim$ to the next level\;
                \If{$victim$ has overlaps in the next level}
                    {Create level for $victim$ to avoid recursion \;}
                }
            }
        }
        }
        
    }

    \tcp{\color{purple}\textbf{Lookup LPA in the \pname{}}}
    \SetKwFunction{FMain}{$lookup$}
    \SetKwProg{Fn}{Function}{:}{}
    \Fn{\FMain{$lpa$}}{
        \ForEach{$ level \in groups[lpa\ mod\ 256] $}
        {
            $seg\_pos = binary\_search(level, lpa)$ \;
            $segment = level.get\_segment(seg\_pos) $ \;
            {\If{$has\_lpa(segment,\ lpa)$}
                {\textbf{return} $ segment.translatePPA(lpa) $ \;}
            }
        }
    }

    \tcp{\color{purple}\textbf{\pname{} Compaction}}
    \SetKwFunction{FMain}{$seg\_compact$}
    \SetKwProg{Fn}{Function}{:}{}
    \Fn{\FMain{}}{
        \ForEach{$group \in groups$}{
            \ForEach{$upper\_level, lower\_level \in group$}{
                \ForEach{$segment \in upper\_level$}{
                    $seg\_update(segment, lower\_level)$
                }
                \If{$upper\_level$ is empty}{
                    $group.remove(upper\_level)$
                }
            }
        }
    }
    \caption{{\sc \pname{} operations}}
    \label{algo:logplr}
    \end{algorithm}

\begin{algorithm}[t]
    \DontPrintSemicolon 
    \SetAlgoNoEnd
    
    \tcp{\color{purple}\textbf{Check if Segment Contains LPA}}
    \SetKwFunction{FMain}{$has\_lpa$}
    \SetKwProg{Fn}{Function}{:}{}
    \Fn{\FMain{$seg,\ lpa$}}{
        $acc \leftarrow seg.accurate$ \;
            {\If{$lpa \not\in [seg.S_{LPA}, seg.S_{LPA} + seg.L]
            \ or \newline
            \quad(not\ acc\ \&\ check(CRB)\ failed)
            \ or \newline
            \quad(acc\ \&\ (lpa - seg.S_{LPA})\ mod\ \lceil \frac{1}{seg.K} \rceil \neq 0)$}
                {$return\ False$ \;}
            }
            $return\ True$ \;
    }
    
    \tcp{\color{purple}\textbf{Convert Segment into a Temporary Bitmap}}
    \SetKwFunction{FMain}{$get\_bitmap$}
    \SetKwProg{Fn}{Function}{:}{}
    \Fn{\FMain{$seg,\ start,\ end$}}{
        $bm \leftarrow bitmap\ of\ length\ (end-start+1)$ \;
        {\ForEach{$lpa \in [start, end]$}{
            {\If{$has\_lpa(seg,\ lpa)$}
                {$bm[lpa-start] = 1$ \;}
            }
            {\Else{$bm[lpa-start] = 0$ \;}
            }
        }
        }
        {\textbf{return} $ bm $ \;}
    }

    \tcp{\color{purple}\textbf{Merge a New Segment with an Old Segment}}
    \SetKwFunction{FMain}{$seg\_merge$}
    \SetKwProg{Fn}{Function}{:}{}
    \Fn{\FMain{$new,\ old$}}{
        $start \leftarrow min(new.S_{LPA},\ old.S_{LPA})$ \;
        $end \leftarrow max(new.S_{LPA}+new.L,\ old.S_{LPA}+old.L)$ \;
        $bm_{new} \leftarrow get\_bitmap(new,\ start,\ end)$ \;
        $bm_{old} \leftarrow get\_bitmap(old,\ start,\ end)$ \;
        $bm_{old} \leftarrow bm_{old}\ \&\ \neg bm_{new} $ \;
        $first,\ last \leftarrow$ the first and last valid bit of $bm_{old}$ \;
        $old.S_{LPA},\ old.L \leftarrow first + start,\ last - first$ \;
        {\If{no valid bits in $old$}{
            {$old.L \leftarrow -1$ \tcp*{mark it as removable} }
        }}
        {\If{$not\ old.accurate$}{
            {Remove outdated LPAs in CRB \;}
        }}
    }
    
    \caption{{\sc Segment Merge}}
    \label{algo:merge}
    \end{algorithm}
    

\noindent
\paragraph{{Creation of Learned Segments.}} {Once the data buffer of the SSD controller is filled, \mbox{\pname{}} takes 
the LPAs and PPAs of the flash pages in the buffer as the input. It sorts the LPA-PPA mappings by reordering the flash pages 
with their LPAs (see $\S$\mbox{\ref{subsec:learnedoptimization}}), 
and uses greedy piecewise linear regression~\mbox{\cite{pgmindex:vldb2014}} to learn the index segment. }

\noindent
\paragraph{Insert/Update of Learned Segments.}
When we insert or update a new learned index segment, we will place it in the topmost level of the log-structured 
mapping table. Since each level of the mapping table is sorted, we can quickly identify its insert location via a 
binary search (line 2 in Algorithm~\ref{algo:logplr}). If the new segment is approximate, \pname{} will 
update the CRB for future lookups (line 4-7 in Algorithm~\ref{algo:logplr}). After that, \pname{} will check whether 
the new segment overlaps with existing segments. If yes, \pname{} will identify the overlapped LPAs.  
The overlap detection is performed by the comparison between the LPA range of the new segment and 
$[S_{LPA}, S_{LPA}+L]$ of the adjacent segments. 
We group these overlapping segments as a list of victim segments (line 8 in Algorithm~\ref{algo:logplr}). 
\pname{} will merge segments to remove outdated LPAs (line 10 in Algorithm~\ref{algo:logplr} and line 14-25 in Algorithm~\ref{algo:merge}).  

To fulfill the segment merge, \pname{} will use the $S_{LPA}$, $L$, and $K$ to reconstruct the list
of the encoded LPAs in the victim segment. And it will create a bitmap to index these encoded LPAs (line 6-13 in Algorithm~\ref{algo:merge}). 
Given an accurate segment with $S_{LPA} = 100, K = 0.5, L = 6$, we can infer that its encoded LPAs 
are $[100, 102, 104, 106]$. We can transfer the LPA list to the bitmap $[1010101]$. 
If the victim segment is an approximate segment, 
\pname{} will leverage the $S_{LPA}$, $L$, and the LPAs stored in the CRB to reconstruct the encoded LPAs. 
Afterwards, 
\pname{} will conduct a comparison between the bitmaps to identify the 
overlapped LPAs (line 15-19 in Algorithm~\ref{algo:merge}).   

\begin{figure}[t]
    \centering
    \includegraphics[width=0.48\textwidth]{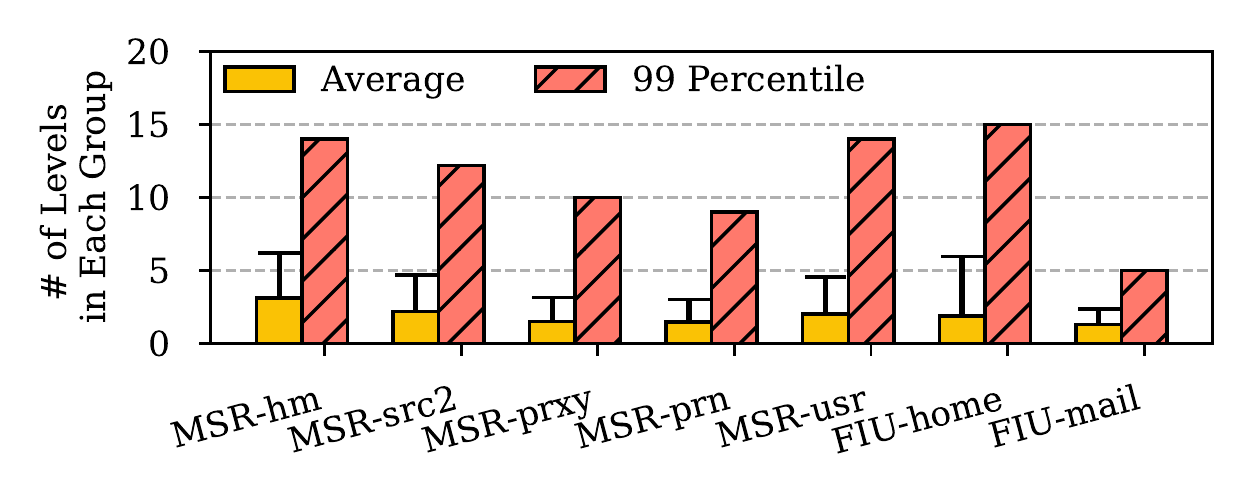}
\vspace{-5ex}
    \caption{A study of the number of levels in the log-structured mapping table for different storage workloads.}
    \label{fig:level}
\end{figure}

\begin{figure}[t]
    \centering
    \includegraphics[width=0.48\textwidth]{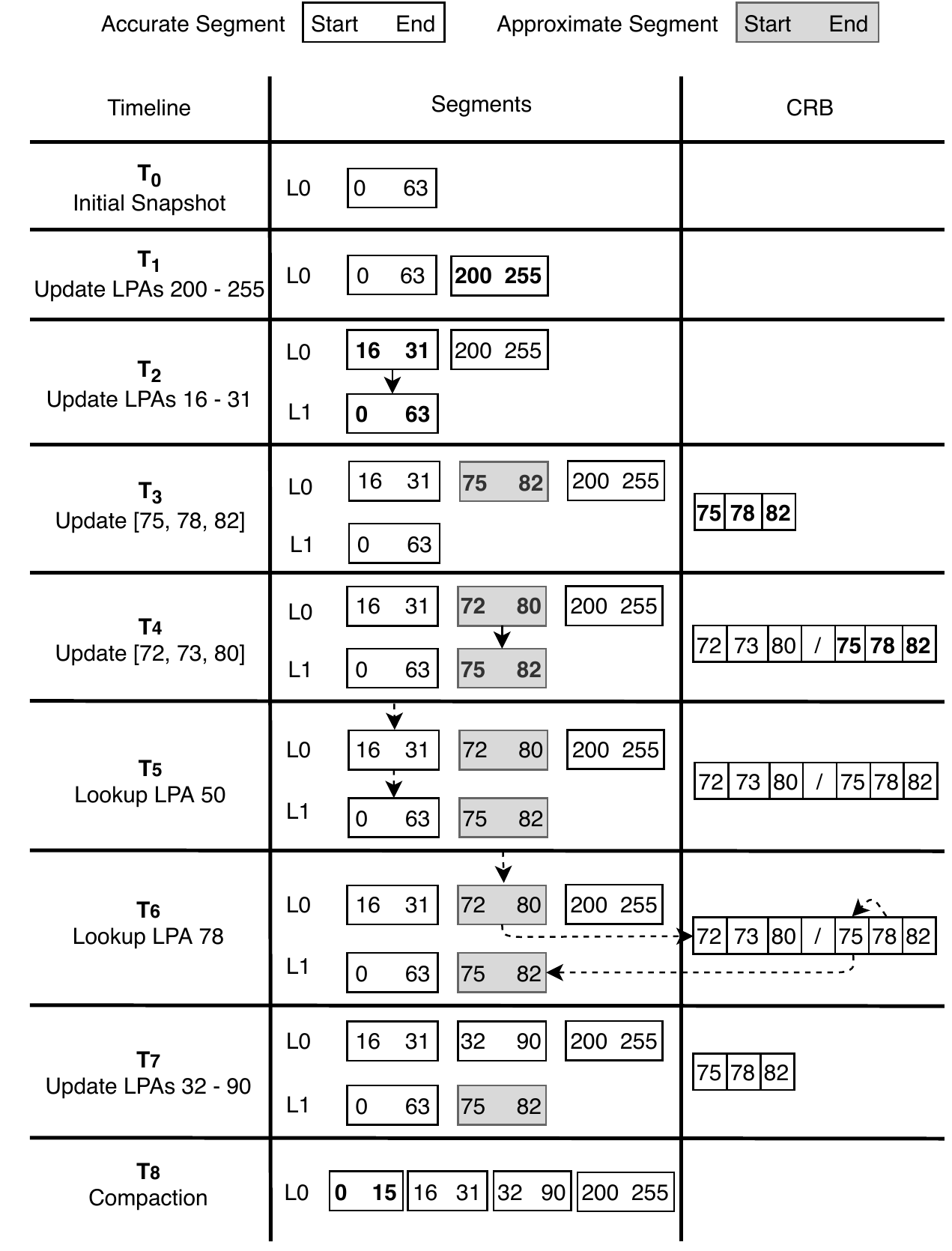}
	\vspace{-4ex}
    \caption{Examples that involve update/insert, lookup, and compaction operations in \pname{}.}
    \label{fig:merge}
	\vspace{-3ex}
\end{figure}

During the segment merge, \pname{} will update the $S_{LPA}$ and $L$ of the old segments accordingly, remove the outdated LPAs from CRB for approximate segments. 
Note that we do not update the $K$ and $I$ for the victim segments during the merge. 

After the merge, (1) if the victim segment does not contain any 
valid LPA ($L$ is negative), it will be removed from the mapping table (line 11-12 in Algorithm~\ref{algo:logplr}).
(2) If the victim segment has valid LPAs but their range still overlaps with the new segment, 
the victim segment will be moved to the next level in the log-structured mapping table (line 13-16 in Algorithm~\ref{algo:logplr}). To avoid recursive updates across the levels, we create a new level for the victim segment if it also overlaps with segments in the next level. According to 
our study of diverse workloads, this will not create many levels in the mapping table (see Figure~\ref{fig:level}). 
(3) If the victim segment has valid LPAs and they do not overlap with the new segment,
we do not need to perform further operations. 
This is because the victim segment is updated with new $S_{LPA}$ and $L$ during segment merge (line 20-25 in Algorithm~\ref{algo:merge}), and the new segment insertion keeps each level sorted (line 3 in Algorithm~\ref{algo:logplr}). 

To facilitate our discussion, we present a few examples in Figure~\ref{fig:merge}. 
At the initial stage, the mapping table has one segment that indexes the LPA range [0, 63]. 
At $T_1$, the new segment [200, 255] is directly inserted into the topmost level, as 
it does not overlap with existing segments. 
At $T_2$, we insert a new segment [16, 31] that has overlaps with 
the old segment [0, 63], \pname{} conducts the segment merge procedure. After that, 
the old segment still has valid LPAs. Thus, it moves to level 1.   
At $T_3$ and $T_4$, we insert two approximate segments [75, 82] and [72, 80], \pname{} will also insert their 
encoded LPAs into the CRB. The segment [75, 82] will be moved to the next level as it overlaps with the new segment [72, 80].

\paragraph{LPA Lookup.}
\pname{} conducts an LPA lookup from the topmost level of the mapping table with binary searches (line 19 in Algorithm~\ref{algo:logplr}). 
We will check whether the LPA is represented by the matched segment (line 21 in Algorithm~\ref{algo:logplr}, line 1-5 in Algorithm~\ref{algo:merge}).
If the $LPA \in [S_{LPA}, S_{LPA}+L]$ of the segment, \pname{} will check the least bit of its $K$. 
If the least bit of $K$ is 0, it is an accurate segment, and \pname{} will use $f(LPA) = \lceil K*LPA + I \rceil$ to 
get the accurate PPA (see $\S$\ref{subsec:learnedsegment}). Otherwise, it is an approximate segment. 
\pname{} will check the CRB to identify the $S_{LPA}$ of the segment, 
following the approach described in Figure~\ref{fig:crb} and $\S$\ref{subsec:learnedmapping}. 
\pname{} will use the same $f(LPA)$ formula to obtain the PPA. 
If the LPA is not found in the top level of the mapping table, \pname{} will 
search the lower levels until a segment is identified.  

We use Figure~\ref{fig:merge} to illustrate the lookup procedure. 
At $T_5$, we conduct the address translation for $LPA = 50$. However, none of the segments 
in the level 0 covers this LPA, \pname{} will continue the search in the level 1 and find 
the accurate segment [0, 63]. 
At $T_6$, we do the address translation for $LPA = 78$. \pname{} finds that the LPA 78 is in the LPA range of 
the segment [72, 80]. Since this is an approximate segment, \pname{} checks the CRB and finds this LPA is actually indexed 
by the segment [75, 82].  
 
With the PPA, \pname{} will read the corresponding flash page and use the reversed mapping (its corresponding 
LPA) in its OOB to verify the correctness of the address translation. 
Upon mispredictions, we will use the approach discussed in $\S$\ref{subsec:error} to handle it. 

\paragraph{Segment Compaction.}
The purpose of the compaction is to merge segments with overlapped LPAs across different levels, 
which further saves memory space. 
\pname{} will iteratively move the upper-level segments into the lower level, until the mapping 
table is fully compacted (line 27 in Algorithm~\ref{algo:logplr}). When an approximate segment is removed, its corresponding CRB entries will also 
be deleted.
As shown in $T_7$ of Figure~\ref{fig:merge}, we insert a new segment [32, 90] which fully covers the LPA range of the segment [72, 80]. After merge, \pname{} removes the old segment [72, 80].
However, some segments in the level 0 still overlap with the segments in the level 1. After $T_8$, \pname{} will remove outdated segments and LPAs.  

\pname{} performs segment compaction after each 1 million writes by default. According to our experiments 
with various storage workloads, the segment compaction of the entire mapping table 
will take 4.1 milliseconds (the time of 20-40 flash writes) on average. 
Consider the low frequency (i.e., once per 1 million writes), 
the compaction incurs trivial performance overhead to storage operations. 


\begin{figure}[t]
    \centering
    \includegraphics[width=0.38\textwidth]{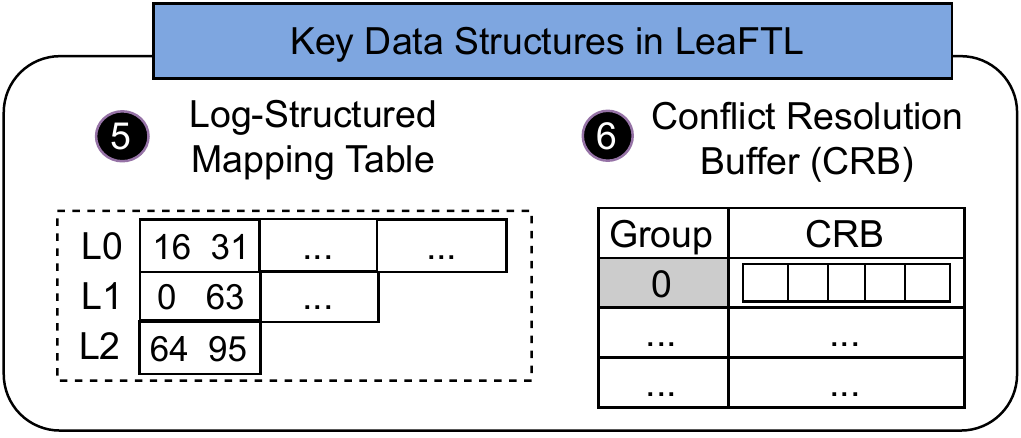}
	\caption{Key data structures used in \pname{}.}
    \label{fig:leaftl_structures}
	\vspace{-2ex}
\end{figure}

\subsection{Put It All Together}
\label{subsec:workflow}
\pname{} is compatible with existing FTL implementations.
As shown in Figure~\ref{fig:leaftl_structures}, it uses the log-structured mapping table 
(\circleb{5}) to replace the address mapping cache (\circleb{1} in Figure~\ref{fig:ftl-structure}), 
and employs CRB (\circleb{6}) for assisting the address translation of approximate segments. 
The CRB requires trivial storage space in the SSD DRAM (see Figure~\ref{fig:crbsize}). 

\paragraph{Read Operation.}
For a read request, \pname{} will first check the data cache. 
For a cache hit, \pname{} serves the read request with the cached flash page.
Otherwise, \pname{} will perform address translation with \circleb{5} (see $\S$\ref{subsec:operation}). 
If there is a misprediction of PPA, 
\pname{} checks the OOB of the mispredicted flash page, 
read the correct page ($\S$\ref{subsec:error}), and updates 
the data cache with the page. 

\paragraph{Write Operation.}
For a write request, \pname{} buffers it in the data cache. 
Once the buffered writes reach the size of a flash block, \pname{} will allocate 
a free block. {It will sort the writes in the buffer based on their LPAs, and 
learn new index segments with the PPAs of the allocated flash block.} This  
enables \pname{} to group more LPA-PPA mappings in the same index segment.  
After that, \pname{} will insert the new index segment in the mapping table, and flush the buffered 
data to the flash blocks. 
For those writes, \pname{} will also check whether their LPAs exist in the mapping table. 
If yes, \pname{} will update their corresponding entries in \circleb{3} BVC and \circleb{4} PVT to indicate 
that they become invalid and can be garbage collected in the future. Otherwise, the new learned  
segments will have their LPA-PPA mappings for future address translations. 

\pname{} caches the mapping table in SSD DRAM for fast lookup. The table will also 
be stored in the flash blocks. \pname{} utilizes the existing \circleb{2} GMD to index 
the translation pages. 
If a segment is not found in the cached mapping table, \pname{} will fetch it 
from the translation blocks and place it in the cached mapping table. 

\noindent
\textbf{{Crash Consistency and Recovery.}}
{Upon system crashes or power failures, \mbox{\pname{}} guarantees the crash consistency of learned indexes.
In order to ensure the data durability of DRAM buffer in SSD controllers, 
modern SSDs today have employed battery-backed DRAM and power 
loss protection mechanisms~\cite{ssd-capacitor, ssd-power}. 
With battery-backed DRAM, 
{\pname{}} has sufficient time to persist the up-to-date mapping table 
to the flash blocks and record their PPAs in the GMD ({\circleb{2}} in Figure~{\ref{fig:ftl-structure}}). During the data recovery, 
\mbox{\pname{}} reads the GMD to locate its mapping table and place it into the DRAM.}

{Without battery-backed DRAM, 
{\pname{}} periodically flushes the learned mapping table and the Block Validity Counter ({\circleb{3}} BVC in Figure~{\ref{fig:ftl-structure})} 
into the flash blocks.
When GC is triggered, {\pname{}} also flushes the updated mapping table and BVC into the flash blocks.  
Upon crashes, {\pname{}} will scan all the flash blocks at the channel-level parallelism, and reconstruct an up-to-date BVC. 
{\pname{}} is able to identify the flash blocks allocated since the last mapping table flush, 
by comparing the up-to-date BVC with the stored BVC in the SSD. Therefore, 
{\pname{}} only needs to relearn the index segments for these recently allocated flash blocks and add them into the mapping table (see $\S$\mbox{\ref{subsec:learnedmapping}}).}


            

\subsection{Implementation Details}
\label{subsec:implt}
\noindent
\textbf{{SSD Simulator.}}
We implement \pname{} based on a trace-driven simulator WiscSim \cite{wiscsim:eurosys2017}, which 
has provided an event simulation environment for the end-to-end performance analysis of SSDs. 
{We extend WiscSim by implementing an LRU-based read-write cache.} \pname{} also preserves the functions of existing FTL, such as GC and wear-leveling. 
To support the learned indexing, 
\pname{} employs a simple linear regression algorithm \cite{plr_algorithm}, which incurs negligible 
computation overhead with modern storage processors (see $\S$\ref{subsec:efficiency}). The error bound $\gamma$ for learned segments is configurable, 
and we set it to 0 by default in \pname{}.  

\noindent
\textbf{{SSD Prototype.}}
{We also develop a real system prototype with an open-channel SSD to validate the functions and efficiency of \mbox{\pname{}}. 
The SSD has 1TB storage capacity with 16 KB flash page size. It has 16 channels, each 
channel has 16K flash blocks, and each flash block has 256 pages. It enables developers to implement their own FTL in the host by providing basic I/O commands such as read, write, 
and erase. We implement \mbox{\pname{}} with 4,016 lines of code using C programming language with the SDK library of the device. 
}


%% file: eval.tex
\section{Evaluation}
\label{sec:eval}
{Our evaluation shows that: (1) {\pname{}} significantly reduces the address mapping table size, 
and the saved memory brings performance benefits ($\S${\ref{subsec:simulation}});
(2) the benefits of \mbox{\pname{}} are validated on a real SSD device ($\S${\ref{subsec:real_device}});
(3) {\pname{}} can achieve additional memory savings and performance benefits with larger error-tolerance, and it demonstrate generality for different SSD configurations ($\S${\ref{subsec:sensitivity}});
(4) Its learning procedure does not introduce much extra overhead to the SSD controller ($\S${\ref{subsec:efficiency}});
(5) It has minimal negative impact on the SSD lifetime ($\S${\ref{subsec:lifetime}}).}

\begin{table}[t]
\centering
\caption{SSD configurations in our simulator.}
\footnotesize 
\begin{adjustbox}{width=.45\textwidth}
	\begin{tabular}{|c|c|c|c|}
\hline
		\textbf{Parameter} & \textbf{Value} & \textbf{Parameter} & \textbf{Value} \\ \hline
		Capacity & 2TB & \#Channels & 16 \\ \hline
		Page size & 4KB & OOB size & 128B \\ \hline
		DRAM size & 1GB & Pages/block & 256 \\ \hline
		Read latency & 20$\mu$s & Write latency & 200$\mu$s \\ \hline
		Erase & 1.5 millisecs & Overprovisioning ratio & 20\%\\ \hline
\end{tabular}
\end{adjustbox}
\label{tab:ssd_ftl}
\end{table}

\begin{table}[t]
    \centering
	\caption{{Real workloads used in our real SSD evaluation.}}
    \footnotesize
    \resizebox{.48\textwidth}{!}{
        \begin{tabular}{|l|l|l|l|}
    \hline
            \textbf{Workload} & \textbf{Description}  \\ \hline
            OLTP~\cite{filebench} &  Transactional benchmark in the FileBench.\\ \hline
            CompFlow (CompF)~\cite{filebench} & File accesses in a computation flow. \\ \hline
            TPCC~\cite{benchbase}& Online transaction queries in warehouses. \\ \hline
            AuctionMark (AMark)~\cite{benchbase} & Activity queries in an auction site. \\ \hline
            SEATS~\cite{benchbase} & Airline ticketing system queries. \\ \hline
    \end{tabular}
    }
    \label{tab:ssd_workload}
    \end{table}

\subsection{Experiment Setup}
\label{subsec:setup}
{We examine the efficiency of {\pname{}} with both the SSD simulator and real SSD prototype.  
As for the evaluation with the SSD simulator, we configure a 2TB SSD with 4KB flash pages and 
1GB DRAM in the SSD controller. We list the core SSD parameters in Table{~\ref{tab:ssd_ftl}}. 
For other parameters, we use the default setting in the WiscSim. We use a variety of storage 
workloads that include the block I/O traces from enterprise servers from Microsoft Research 
Cambridge~{\cite{MSR}} and workload traces from computers at FIU~{\cite{FIU}}. 
As for the evaluation with the real SSD prototype (see $\S$\mbox{\ref{subsec:implt}}), 
we validate the benefits of {\pname{}} using a set of real-world file system benchmarks 
and data intensive applications as shown in Table~\mbox{\ref{tab:ssd_workload}}. 
Before we measure the performance,
we run a set of workloads consisting of various real-world and synthetic storage workload traces to warm up the SSD 
and make sure the GC will be executed during the experiments. 
}

We compare {\pname{}} with state-of-the-art page-level mapping schemes described as follows 
\footnote{We do not compare \pname{} with block-level and hybrid-level mappings, as they perform 
dramatically worse than the page-level mapping~\mbox{\cite{DFTL, sftl:msst2011}}.}.




\begin{itemize}[leftmargin=*]

\vspace{0.5ex}
	\item \textbf{DFTL (Demand-based FTL)}~\cite{DFTL}: 
		it uses a page-level mapping scheme, and  
		caches the most recently used address translation entries in the SSD DRAM. 

\vspace{0.5ex}
\item 
	\textbf{SFTL (Spatial-locality-aware FTL)}~\cite{sftl:msst2011}: it is a page-level mapping that 
		exploits the spatial locality and strictly sequential access patterns of workloads to condense mapping table entries. 

\end{itemize}


\begin{figure}[t]
    \centering
    \includegraphics[width=0.45\textwidth]{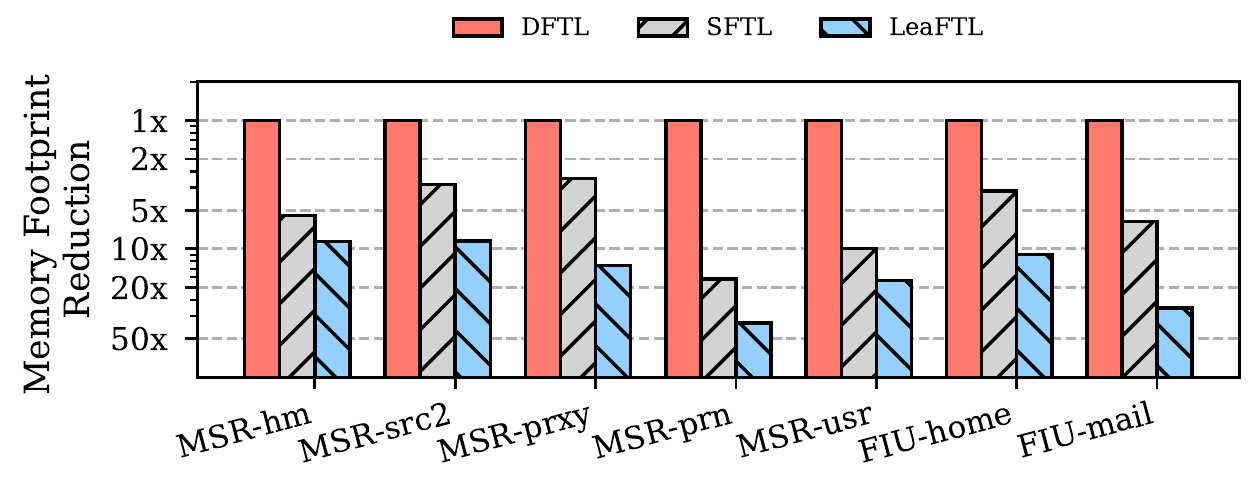}
	\vspace{-4ex}
	\caption{The reduction on the mapping table size of \pname{}, in comparison with DFTL and SFTL.}
	\vspace{-1.5ex}
    \label{fig:memory}
\end{figure}

\begin{figure}[t]
    \centering
    \captionsetup[subfloat]{skip=-1ex}
    \subfloat[SSD performance when using its DRAM mainly for the address mapping table (lower is better).]{
        \centering
        \includegraphics[width=0.48\textwidth]{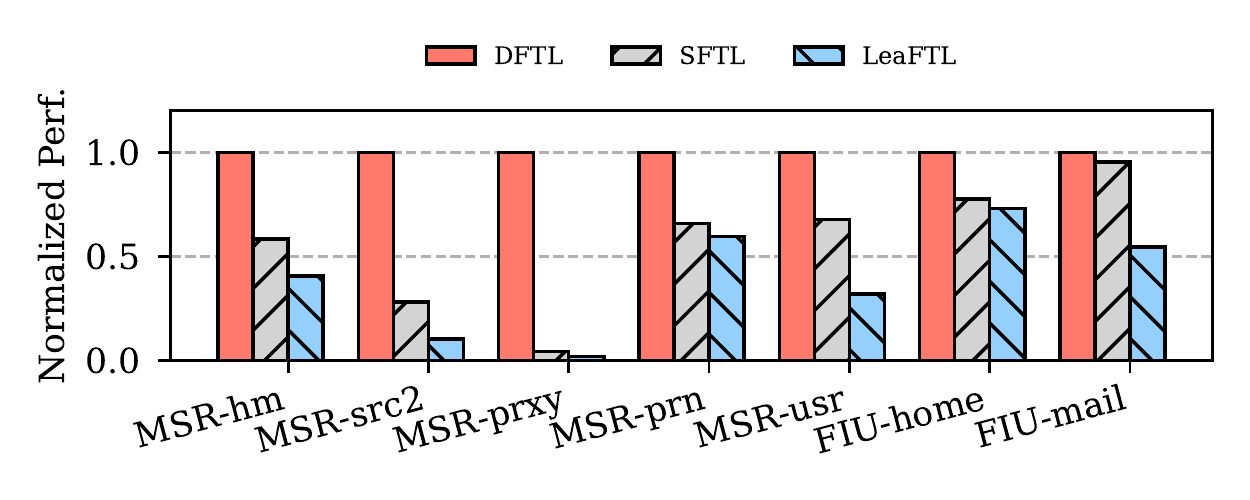}
        \label{fig:workload_latency}
    }
    \hfill
    \subfloat[SSD performance when using its DRAM partially (up to 80\%) for the address mapping table (lower is better).]{
        \centering
        \includegraphics[width=0.48\textwidth]{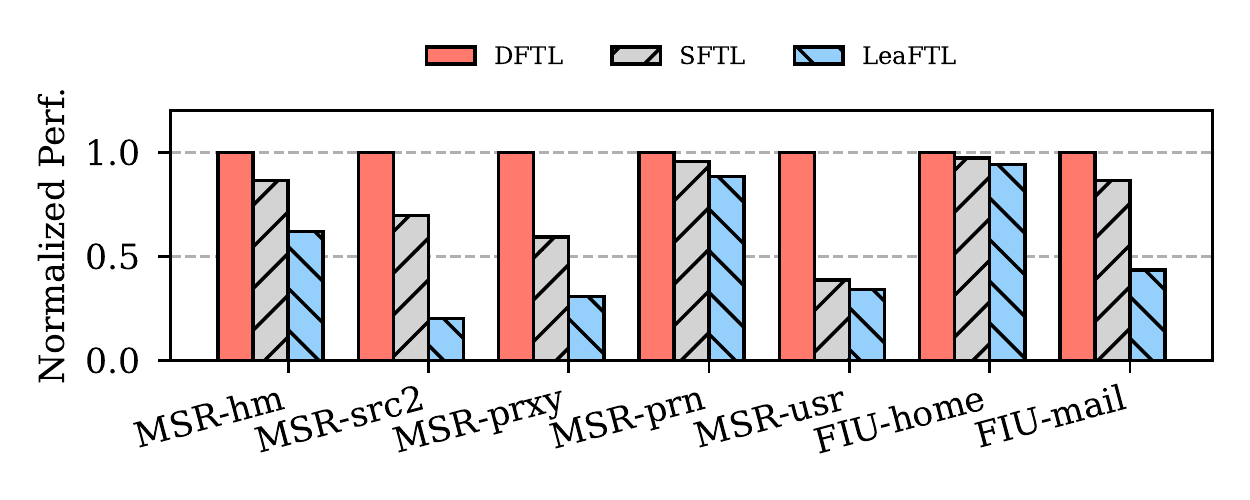}
        \label{fig:workload_latency}
    }
	\caption{Performance improvement with \pname{}.}
    \label{fig:latency}
	\vspace{-1ex}
\end{figure}



\subsection{Memory Saving and Performance}
\label{subsec:simulation}
{We first evaluate the benefits of \mbox{\pname{}} on the memory saving and storage performance with the SSD simulator.}
As shown in Figure~\ref{fig:memory}, \pname{} reduces the mapping table 
size by 7.5--37.7$\times$, compared to the page-level mapping scheme DFTL. This is because \pname{} 
can group a set of page-level mapping entries into an 8-byte segment. In comparison with SFTL, 
\pname{} achieves up to 5.3$\times$ (2.9$\times$ on average) reduction on the address mapping table 
for different storage workloads, when we set its $\gamma = 0$ (i.e., the learned segments are 
100\% accurate). This is because \pname{} captures more LPA-PPA mapping 
patterns. 

We now evaluate the performance benefit of \pname{} from its saved memory space.
We evaluate \pname{} with two experimental settings: (1) the SSD DRAM is mainly used (as much as possible) 
for the mapping table; (2) the SSD DRAM is partially used for the mapping table, in which 
we ensure at least 20\% of the DRAM will be used for the data caching. 


\begin{figure}[t]
    \centering
    \captionsetup[subfloat]{skip=-1ex}
    \centering
    \includegraphics[width=0.48\textwidth]{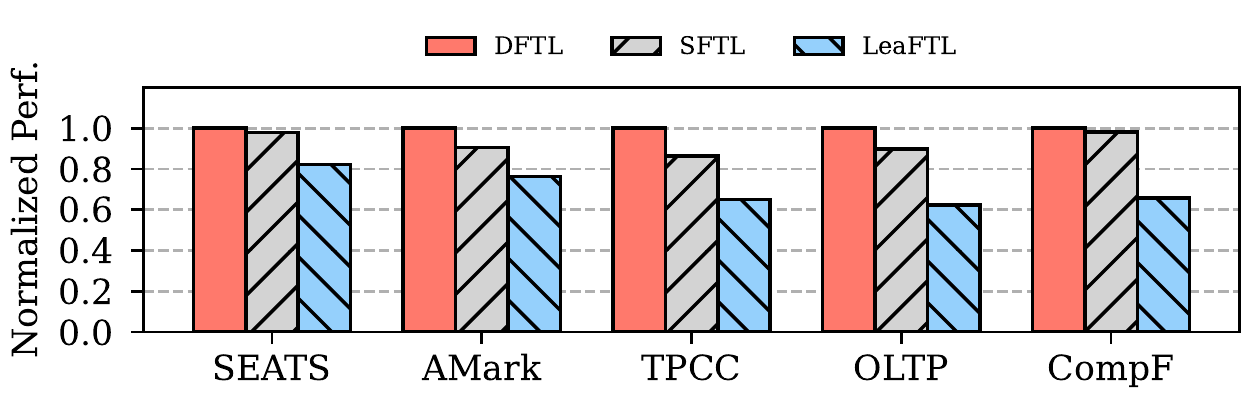}
    \vspace{-3ex}
	\caption{{Performance on the real SSD prototype.}}
     \vspace{-3ex}
    \label{fig:workload_latency}
\end{figure}

\begin{figure}[t]
    \centering
    \captionsetup[subfloat]{skip=-1ex}
    \centering
    \includegraphics[width=0.48\textwidth]{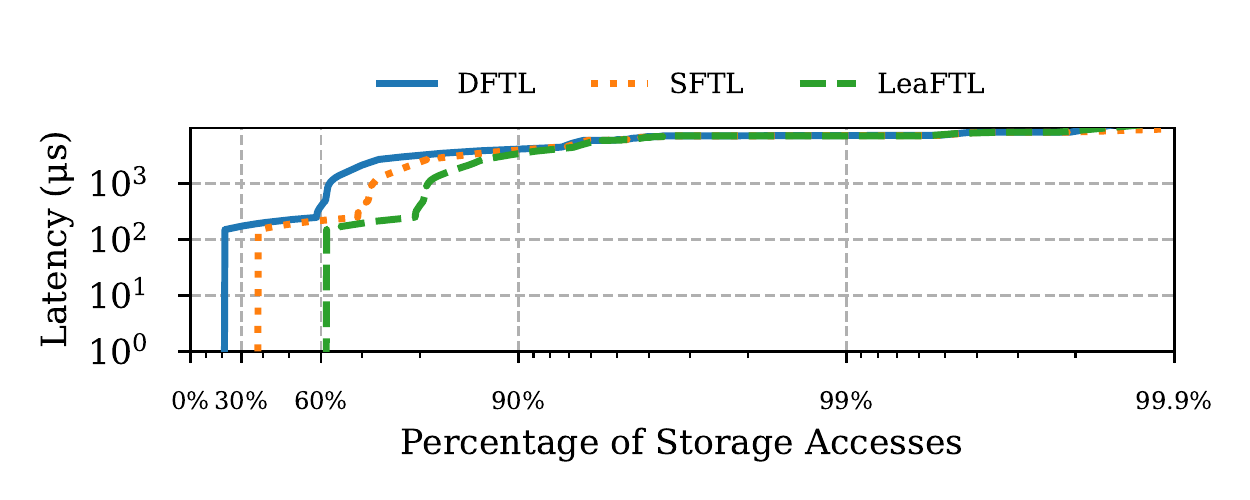}
    \vspace{-5ex}
	\caption{{The latency distribution of storage accesses when running OLTP workload on the real SSD prototype.}}
    \vspace{-3ex}
    \label{fig:workload_latency_cdf}
\end{figure}

In the first setting, DRAM is almost used for mapping table in DFTL.
As shown in Figure~\ref{fig:latency} (a), \pname{} reduces the storage access 
latency by 1.6$\times$ on average (up to 2.7$\times$), compared to SFTL. 
This is because \pname{} saves more memory from the mapping table than SFTL. 
SFTL slightly outperforms DFTL, because it reduces the mapping table size 
by compressing mapping entries with grouping strictly sequential data accesses. 
In the second setting, as shown in Figure~\ref{fig:latency} (b), 
\pname{} obtains 1.4$\times$ (up to 3.4$\times$) and 1.6$\times$ (up to 4.9$\times$) performance speedup, 
compared to SFTL and DFTL, respectively.

\subsection{Benefits on the Real SSD Prototype}
\label{subsec:real_device}

{We validate the benefits of \mbox{\pname{}} on the real SSD prototype with real workloads (see Table~\mbox{\ref{tab:ssd_workload}}). 
They include filesystem benchmark suite FileBench~\mbox{\cite{filebench}}, and transactional database workloads from BenchBase~\mbox{\cite{benchbase, benchbase:control}}. 
All these workloads run on the ext4 file system. With FileBench, we run OLTP and CompFlow (CompF) workloads to read/write 10GB files. 
With BenchBase, we run TPCC, AuctionMark (AMark), and SEATS workloads on MySQL, and their database sizes are 10--30GB. These database 
workloads will generate 37--230GB read traffic and 26--59GB write traffic to the SSD. 
We allocate 256MB DRAM to host the mapping table (for different DRAM sizes, see our sensitivity analysis in $\S${\ref{subsec:sensitivity}}).} 

{We present the performance benefit of \mbox{\pname{}} in Figure~\mbox{\ref{fig:workload_latency}}. 
Across all workloads, \mbox{\pname{}} obtains 1.4$\times$ performance speedup on 
average (up to 1.5$\times$), compared to SFTL and DFTL. Similar to our evaluation with the SSD simulator implementation, the performance benefit 
of \mbox{\pname{}} comes from the memory saving from the address mapping table. 
And \mbox{\pname{}} demonstrates comparable performance improvement on real SSD devices, in comparison with the SSD simulator in $\S$\mbox{\ref{subsec:simulation}}.
We also show the latency distribution of storage accesses in Figure~\mbox{\ref{fig:workload_latency_cdf}}, 
when running the OLTP workload on the real SSD prototype. In comparison with existing FTL schemes, \mbox{\pname{}} does not increase the tail latency of 
storage accesses. And the higher cache hit ratio of \mbox{\pname{}} brings latency reduction for many storage accesses.  
}

\begin{figure}[t]
    \centering
    \includegraphics[width=0.48\textwidth]{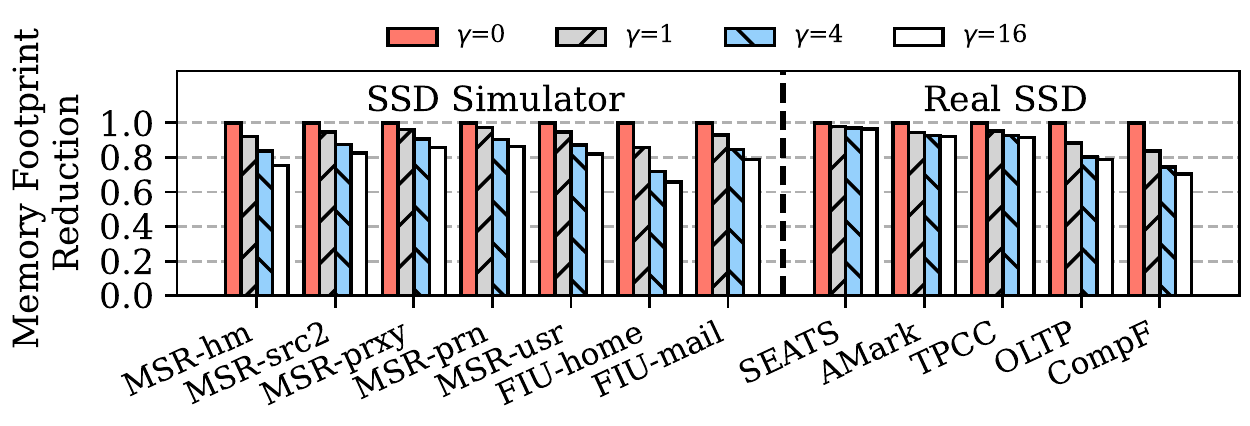}
	\vspace{-4ex}
	\caption{{The reduction of the mapping table size of \mbox{\pname{}} with different $\gamma$ (lower is better).}}
    \label{fig:memory_sensitivity}
\end{figure}

\begin{figure}[t]
    \centering
    \includegraphics[width=0.48\textwidth]{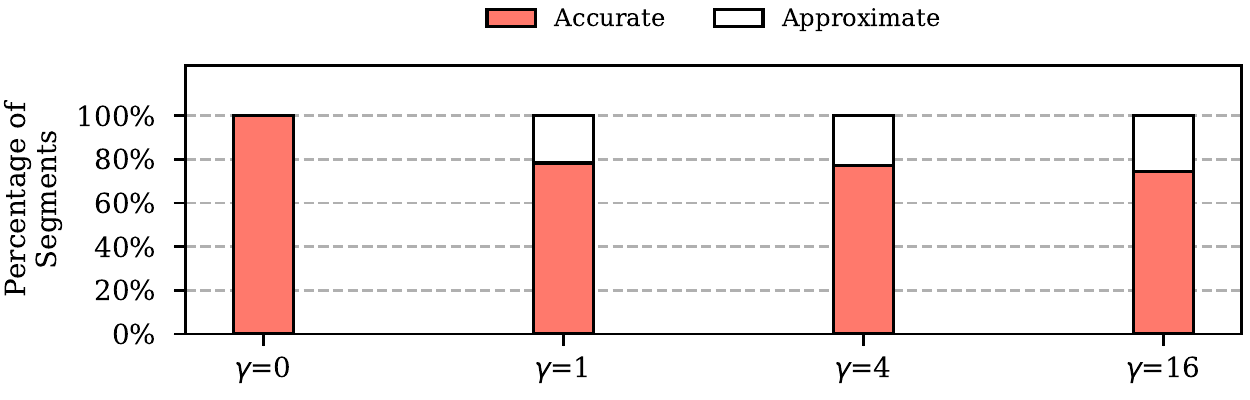}
	\vspace{-4ex}
	\caption{The distribution of learned segments.}
	\vspace{-2ex}
    \label{fig:dist_segments}
\end{figure}

\subsection{Sensitivity Analysis}
\label{subsec:sensitivity}

\begin{figure}[t]
    \centering
    \includegraphics[width=0.48\textwidth]{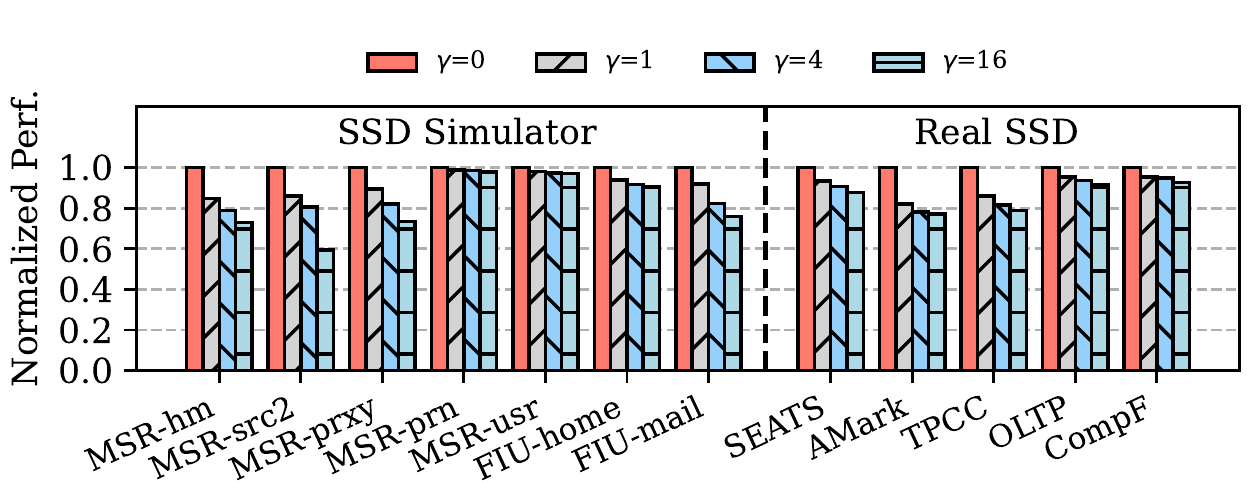}
	\vspace{-4ex}
	\caption{{Performance with various $\gamma$ (lower is better).}}
    \vspace{-1ex}
    \label{fig:latency_sensitivity}
\end{figure}

\noindent
{\textbf{Vary the value of $\gamma$.}
As we increase the value of $\gamma$ from 0 to 16, the size of the learned mapping table is reduced,  
as shown in Figure~{\ref{fig:memory_sensitivity}}.  
{\pname{}} achieves 1.3$\times$ reduction on average (1.2$\times$ on the real SSD) with $\gamma=16$, compared to that of $\gamma=0$. 
The saved memory with a larger $\gamma$ is achieved by learning a wider range of LPAs into approximate segments. 
To further understand this, we profile the distribution of segments learned by {\pname{}} with different values of $\gamma$, 
as shown in Figure~{\ref{fig:dist_segments}}. When $\gamma=0$, all the segments are accurate. When $\gamma=16$, 26.5\% of 
the learned segments are approximate on average, and  
{\pname{}} delivers 1.3$\times$ improvement on storage performance (1.2$\times$ with workloads on the real SSD), in comparison 
with the case of $\gamma=0$ (see Figure~{\ref{fig:latency_sensitivity}}).}

\begin{figure}[t]
    \centering
    \includegraphics[width=0.45\textwidth]{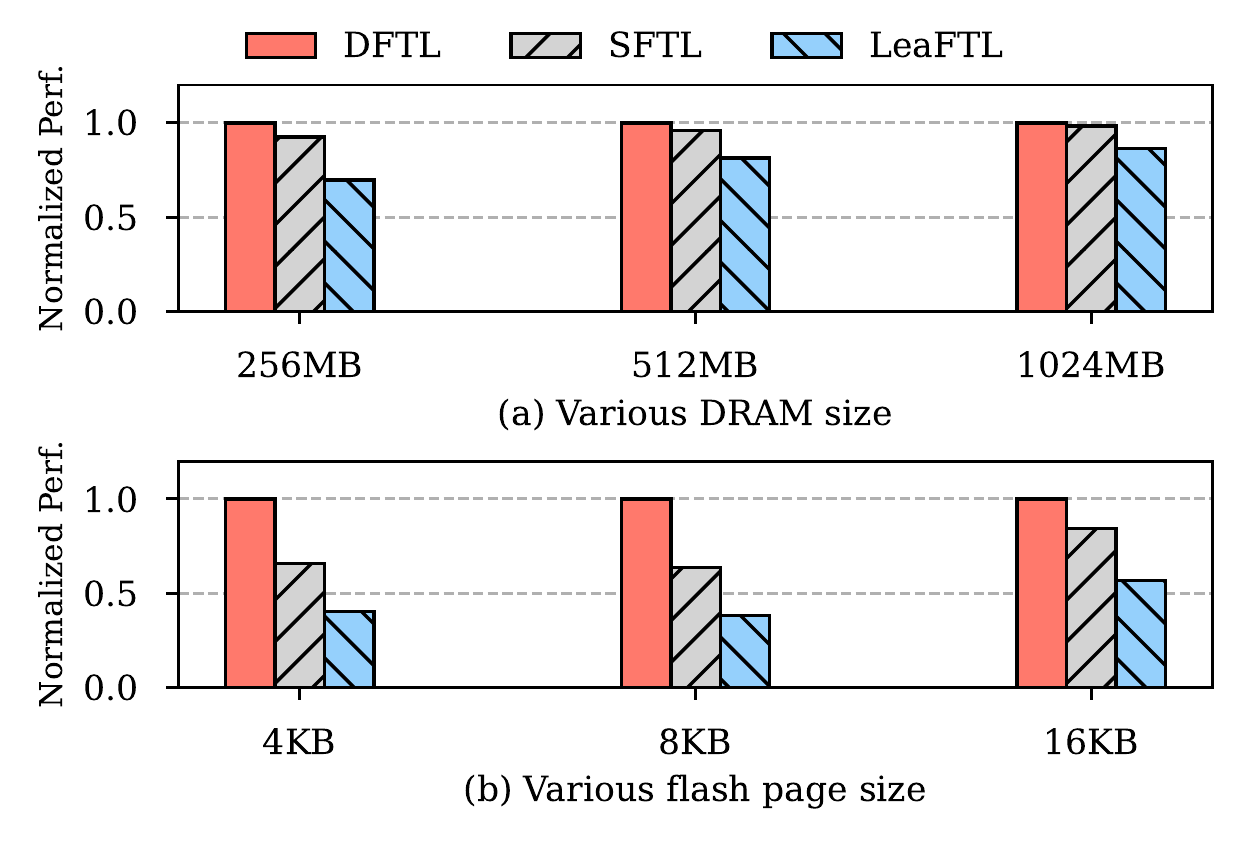}
	\vspace{-2.5ex}
	\caption{{SSD performance with different DRAM capacity and flash page size (lower is better).}}
	\vspace{-1ex}
    \label{fig:page_size_latency}
\end{figure}

\noindent
{\textbf{Vary the SSD DRAM capacity}.
We now conduct the sensitivity analysis of SSD DRAM by varying its capacity from 256MB to 1GB on the real SSD prototype.
As shown in Figure~{\ref{fig:page_size_latency}} (a), {\pname{}} always outperforms DFTL and SFTL as we vary the SSD DRAM capacity. 
As we increase the DRAM capacity, the storage workloads are still bottlenecked by the available memory space for the data caching. 
\mbox{\pname{}} can learn various data access patterns and significantly reduce the address mapping table size, 
the saved memory further benefits data caching. 
}


\noindent
\textbf{Vary the flash page size.}
In this experiment, 
we fix the number of flash pages, and {vary the flash page size from 4KB to 16KB in the SSD simulator}, as SSD vendors usually 
use larger flash pages for increased SSD capacity. {We use the simulator for this study, since the flash page size of the real SSD 
is fixed.} 
As shown in Figure~\ref{fig:page_size_latency} (b), \pname{} always performs the best in comparison with DFTL and SFTL. 
As we increase the flash page size to 16KB, we can cache less number of flash pages with limited DRAM capacity. Thus, \pname{} experiences a slight performance drop. 
As we fix the total SSD capacity and vary the page size, \pname{} outperforms SFTL by 1.2$\times$ and 1.1$\times$ for the page size of 
8KB and 16KB, respectively. 

\subsection{Overhead Source in \pname{}}
\label{subsec:efficiency}
We evaluate the overhead sources in \pname{} in three aspects: 
(1) the performance overhead of the learning procedure in \pname{}; (2) the LPA lookup overhead 
in the learned segments; and (3) the overhead caused by the address misprediction in \pname{}. 



We evaluate the performance of segment learning and address lookup on an ARM Cortex-A72 core. This core is  
similar to the storage processor used in modern SSDs.  
The learning time for a batch of 256 mapping entries is 9.8--10.8 $\mu$s (see Table~\ref{tbl:arm}). 
As we learn one batch of index segments for every 256 flash writes, the learning overhead 
is only 0.02\% of their flash write latency. 

\begin{figure}[t]
    \centering
    \includegraphics[width=0.48\textwidth]{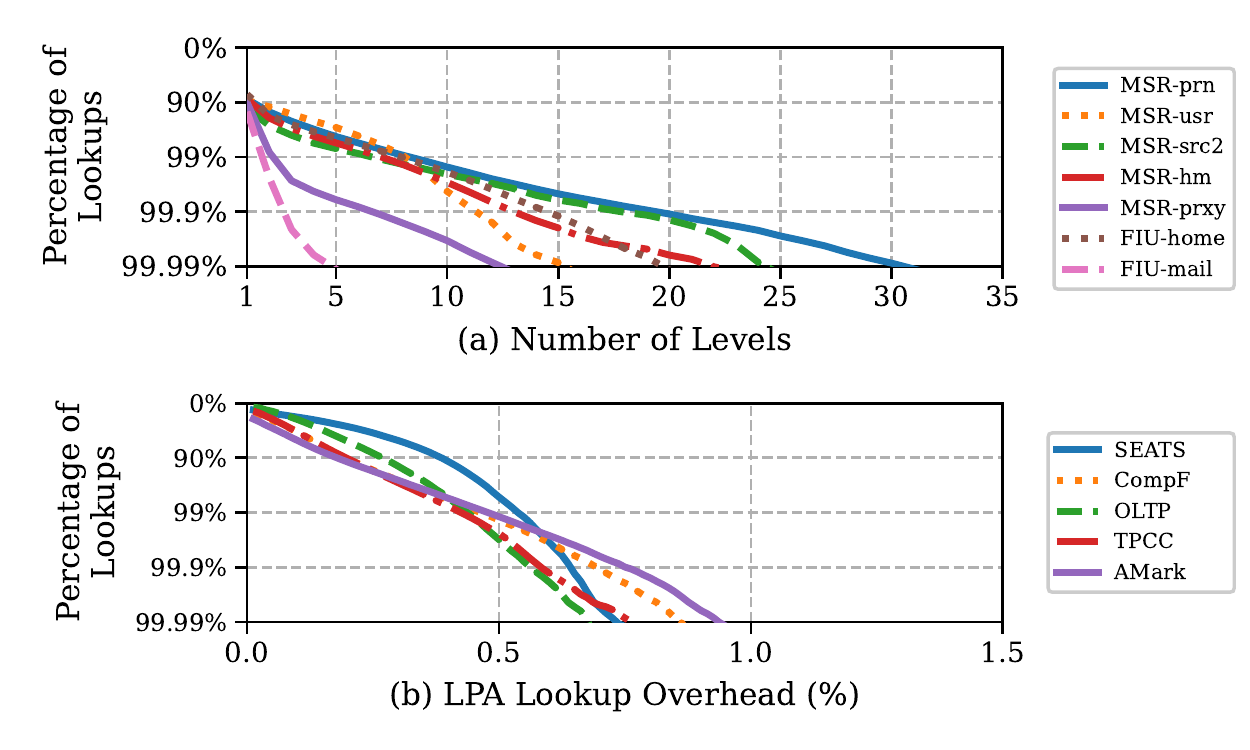}
	\vspace{-5ex}
    \caption{{Performance overhead of the LPA lookup.}}
    \label{fig:lookup_cdf}
\end{figure}

\begin{figure}[t]
    \centering
    \includegraphics[width=0.48\textwidth]{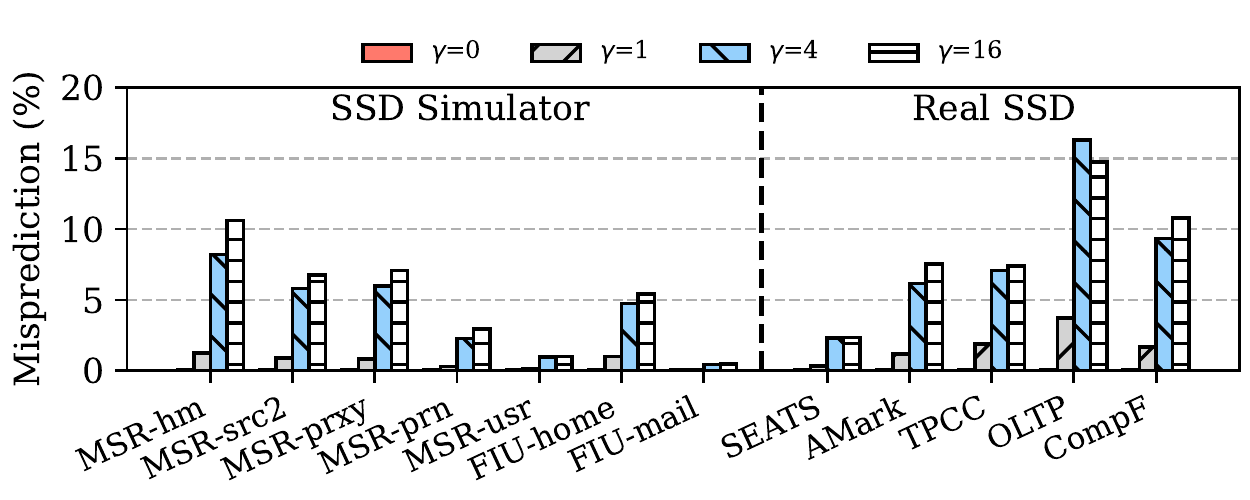}
	\vspace{-4ex}
    \caption{{Misprediction ratio of flash pages access.}}
    \vspace{-1ex}
    \label{fig:misprediction}
\end{figure}

In \pname{}, the LPA lookup is 40.2--67.5 ns, as the binary search of segments is fast and some
segments can be cached in the processor cache. The lookup time is slightly higher as we increase 
$\gamma$, due to the additional CRB accesses. We also profile the cumulative distribution function (CDF) of the number of levels to lookup 
for each LPA lookup, and present the results in Figure~\ref{fig:lookup_cdf} (a). 
For most of the tested workloads, $90\%$ of the mapping table lookup can be fulfilled at the topmost level, 
and $99\%$ of the lookups are within 10 levels. Although MSR-prn workload requires more lookups 
than other workloads, it only checks $1.4$ levels on average. {We also evaluate the performance overhead of the LPA lookup on the real SSD, and show the results 
in Figure~{\ref{fig:lookup_cdf}} (b). The extra lookup overhead for each flash read is $0.21\%$ on average. And for $99.99\%$
of all the lookups, the additional overhead is less than $1\%$ of the flash access latency.
}

\begin{table}[t]\caption{Overhead source of \pname{} with an ARM core.}
 	\vspace{-3ex}
\footnotesize
\centering
\begin{adjustbox}{width=.42\textwidth}
\begin{tabular}{|c|c|c|c|}
\hline
$\gamma$ & 0 & 1 & 4 \\ \hline
Learning (256 LPAs) & 9.8 $\mu$s & 10.8 $\mu$s & 10.8 $\mu$s \\ \hline
Lookup (per LPA) & 40.2 ns & 60.5 ns & 67.5 ns \\ \hline
\end{tabular}
\end{adjustbox}
\label{tbl:arm}
\end{table}

\pname{} also has low misprediction ratios with approximate segments. 
This is because \pname{} can still learn accurate segments even if $\gamma > 0$, 
and not all entries in the approximate segments will result in misprediction. 
{As shown in Figure~{\ref{fig:misprediction}}, most of the workloads achieve less than 10\% misprediction ratio when $\gamma = 16$. 
We obtain similar misprediction ratio on the real SSD prototype.} 
Note that each misprediction only incurs one flash read access with the help of our proposed OOB verification.

\subsection{Impact on SSD Lifetime}
\label{subsec:lifetime}
The flash blocks of an SSD can only undergo a certain amount of writes. 
In this experiment, we use the write amplification factor (WAF, 
the ratio between the actual and requested flash writes) to evaluate the SSD lifetime. 
The SSD will age faster if the WAF is larger. 
As shown Figure~\ref{fig:waf}, the WAF of \pname{} is comparable to DFTL and SFTL. 
DFTL has larger WAF in most workloads. SFTL and \pname{} occasionally flush translation pages 
to the flash blocks, but the cost is negligible.

%% file: discussion.tex
\section{Discussion}
\label{sec:discussion}

\noindent
\textbf{{Why Linear Regression.}}
{Unlike deep neural networks, the linear regression used in \mbox{\pname{}} is simple and lightweight, which takes only 
a few microseconds to learn an index segment with embedded ARM processors available in modern SSD controllers. In addition, 
the linear regression algorithm has been well studied, and offers guaranteed error bounds for its learned results. \mbox{\pname{}} is the 
first work that uses learning techniques to solve a critical system problem (i.e., address mapping) in SSDs.}

\noindent
\textbf{{Adaptivity of \mbox{\pname{}}.}}
{\mbox{\pname{}} focuses on the page-level address translation, its design and implementation will not be affected by the 
low-level flash memory organization (i.e., TLC/QLC). As we use TLC/QLC technique to further increase the SSD capacity, 
the address mapping issue will become more critical, since the SSD DRAM capacity does not scale well and becomes the 
bottleneck for caching address mappings and user data.}

\noindent
\textbf{{Recovery of Learned Index Segments.}}
{As discussed in $\S$\mbox{\ref{subsec:workflow}}, using a battery or large capacitor to preserve and persist the 
cached segments upon failures or crashes will simplify the recovery procedure significantly. In our real 
SSD prototype, we do not assume the battery-backed DRAM is available. Thus, we follow the conventional recovery 
approach in modern SSDs~\mbox{\cite{DFTL, FlashMap}}, and scan flash blocks in parallel by utilizing the channel-level parallelism. 

When we run real workloads like TPCC on the SSD prototype, we intentionally reboot the system after running the workload for a period 
of time (0.5-3 hours). We find that the system can recover in 15.8 minutes on average whenever the reboot happens. This  
is similar to the time of recovering the conventional page-level mapping table in DFTL~\mbox{\cite{DFTL}}. This is mostly caused 
by scanning the blocks in a channel (70MB/s per channel in our SSD prototype), and the time for reconstructing 
recently learned segments is relatively low (101.3 milliseconds on average). We believe the recovery time is not much of a concern as the recovery does not happen frequently 
in reality. And the recovery can be accelerated as we increase the channel-level bandwidth. 
In addition, if an SSD can tolerate more data losses, we can still ensure the crash consistency by only loading the stored index segments from 
flash chips, which requires minimum recovery time. 
}

\begin{figure}[t]
    \centering
	\vspace{-1ex}
    \includegraphics[width=0.48\textwidth]{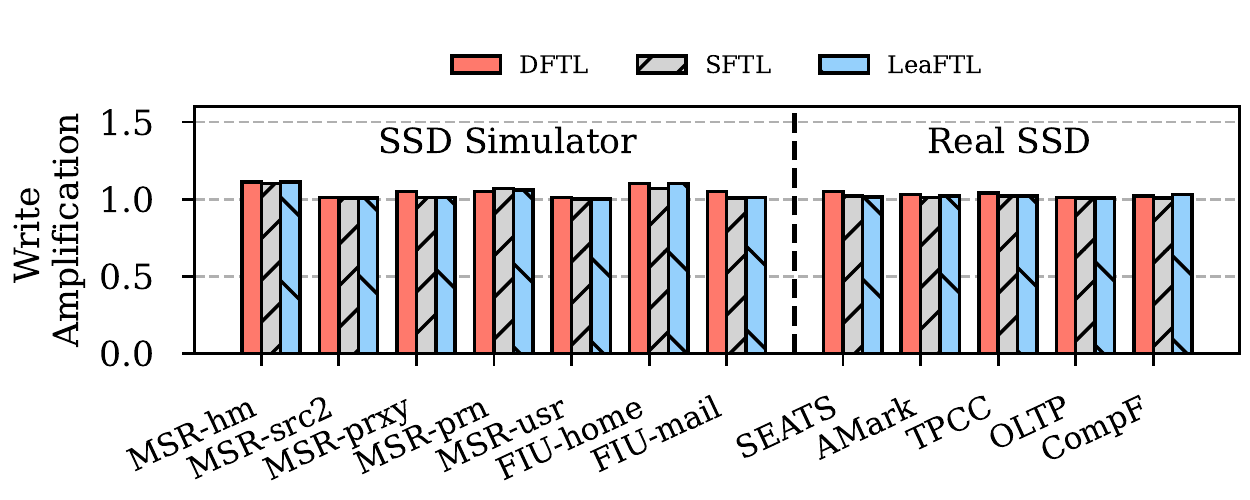}
	\vspace{-4ex}
	\caption{{Write amplification factor of \mbox{\pname{}}.}}
 	\vspace{-4ex}
    \label{fig:waf}
\end{figure}

%% file: related.tex
\section{Related Work}
\label{sec:related}

\noindent
\textbf{Address Translation for SSDs.}
A variety of FTL optimizations have been proposed~\cite{DFTL, 
dayan:sigmod2016, sftl:msst2011, chen2014demand, qin2010demand, jung2010superblock, kwon2010janus, 
park2008reconfigurable}. These works exploited the data locality of flash accesses to 
improve the cache efficiency of the mapping table. 
However, most of them were developed with human-driven heuristics. 
An alternative approach is to integrate application semantics 
into the FTL, such as content-aware FTL~\cite{chen:fast2011}. However, 
they were application specific and required significant changes to the FTL. \pname{} is a generic solution 
and does not require application semantics in its learning.
Researchers proposed to integrate 
the FTL mapping table into the host~\cite{FlashMap, zhang:fast2012, www-directcache, josephson:storage2010}. Typical examples include 
DFS~\cite{josephson:storage2010}, Nameless writes~\cite{zhang:fast2012}, FlashMap~\cite{FlashMap}, and FlatFlash~\cite{flatflash:asplos2019}. 
\pname{} is orthogonal 
to them and can be applied to further reduce their memory footprint.        
 
\noindent
\textbf{Machine Learning for Storage.}
Recent studies have been using learning techniques to build indexes such as B-trees, log-structured 
merge tree, hashmaps, and bloom filters~\cite{learnedindex:sigmod2018, ferr:icml2020,cdfshop:sigmod2020,
radixspline:aidm2020,pgmindex:vldb2020, learnedindex:osdi2020} for in-memory datasets, identify optimal 
cache replacement and prefetching policies~\cite{liu2020imitation,sethumurugan2021designing,
rodriguez2021learning,shi2019applying}, facilitate efficient storage harvesting~\cite{storageharvesting:osdi22}, 
and drive the development of software-defined storage~\cite{lssd:nips22}. 
\pname{} applies learning techniques to optimize the 
address mapping. 
However, 
unlike existing optimizations~\cite{nan:survey2021, 
learnedaddr:nips2018} such as learned page table for virtual memory that used deep neural 
networks to learn the patterns, \pname{} provides a lightweight solution. 



\noindent
\textbf{SSD Hardware Development.}
For the recent SSD 
innovations~\cite{gartner_2017, flashmemory, flashhistory, 3dnand} like
Z-SSD~\cite{znand}, 
KVSSD~\cite{kvssd:samsung}, and ZNS SSD~\cite{zns:osdi2021},  
DRAM capacity and storage processor are still the main constraints in SSD controllers. 
As we scale the storage capacity, 
the challenge with the address 
translation becomes only worse. 
Researchers recently deployed hardware accelerators inside SSD controllers for near-data 
computing~\cite{lee2020smartssd, deepstore:micro2019, insider:atc2019, smartssd}. 
We wish to extend \pname{} with in-storage accelerators to deploy more powerful learning models as the future work.

%% file: conclusion.tex
\section{Conclusion}
\label{sec:conclusion}
We present a learning-based flash translation layer, named \pname{} for SSDs. 
\pname{} can automatically learn different flash access patterns and build space-efficient 
indexes, 
which reduces the address 
mapping size and improves the caching efficiency in the SSD controller. Our evaluation 
shows that \pname{} improves the SSD performance by 1.4$\times$ on average for a variety of 
storage workloads.

%% file: ack.tex
\section*{Acknowledgments}
\label{sec:ack}
We thank the anonymous reviewers for their helpful comments and
feedback. 
This work is partially supported by the NSF CAREER Award 2144796, CCF-1919044, and CNS-1850317.